# Revealing Higher-Order Topological Bulk-boundary Correspondence in Bismuth Crystal with Spin-helical Hinge State Loop and Proximity Superconductivity

Dongming Zhao[1†], Yang Zhong[1†], Tian Yuan[1†], Haitao Wang[1], Tianxing Jiang[1], Yang Qi[1*], Hongjun Xiang[1,3,4], Xingao Gong[1,3,4*], Donglai Feng[2,3,4,5*], Tong Zhang[1,3,4,5*]

[1)] Department of Physics, State Key Laboratory of Surface Physics and Advanced Material Laboratory, Fudan University, Shanghai 200438, China
[2)] New Cornerstone Laboratory, National Synchrotron Radiation Laboratory and School of Nuclear Science and Technology, University of Science and Technology of China; Hefei 230027, China
[3)] Collaborative Innovation Center for Advanced Microstructures, Nanjing 210093, China
[4)] Shanghai Research Center for Quantum Sciences, Shanghai 201315, China
[5)] Hefei National Laboratory, Hefei 230088, China

† These authors contributed equally.
* Corresponding authors

**Topological materials are typically characterized by gapless boundary states originated from nontrivial bulk band topology, known as topological bulk-boundary correspondence. Recently, this fundamental concept has been generalized in higher-order topological insulators (HOTIs). *E.g.*, a second-order three-dimensional (3D) TI hosts one-dimensional (1D) topological hinge states winding around the crystal. However, a complete verification of higher-order topology is still lacking as it requires probing all the crystal boundaries. Here we studied a promising candidate of second-order TI, bismuth (Bi), in the form of mesoscopic crystals grown on superconducting $V_3Si$. Using low-temperature scanning tunneling microscopy, we directly observed dispersive 1D states on various hinges of the crystal. Upon introducing magnetic scatterers, new scattering channels emerged selectively on certain hinges, revealing their spin-helical nature. Combining first-principle calculation and global symmetry analysis, we find these hinge states are topological and formed a closed loop encircling the crystal. This provides direct evidence on the higher-order topology in Bi. Moreover, proximity superconductivity is observed in the topological hinge states, enabling HOTI as a promising platform for realizing topological superconductivity and Majorana quasiparticles.**

Topology is a mathematical concept describing the global property of geometric objects [1]. For example, the surfaces of a sphere and a torus could locally appear similar [Fig. 1(a)], but they have distinct global topological quantities (genus of 0 and 1, respectively). In the past decade, topology in electronic structure of solids has been intensively explored [2-6], leading to the discovery of topologically nontrivial materials like three-dimensional (3D) topological insulators (TI) and superconductors (TSC) [7,8]. The defining property of these materials is that besides the gapped bulk, there always exist metallic states at *all* the surfaces which are protected by symmetry [Fig. 1(b)], known as topological bulk-boundary correspondence [8]. In principle, a complete verification of such global topological property requires the examination of all the boundaries, which has never been carried out for any practical materials. Even for the well-known 3D TIs $Bi_2Se_3$ and $Bi_2Te_3$ [7-9], their topological surface states were only observed on (111) surface.

Recently, a new class of topological material called higher-order topological insulators (HOTIs) has been proposed, which generalizes the bulk-boundary correspondence principle [10-16]. Different from the (first-order) 3D TI, second-order 3D TIs possess gapped bulk and surface, but exhibit 1D gapless states on specific hinges that protected by symmetries [Fig. 1(c)]. These topological hinge states can be spin-helical or chiral (depends on whether time-reversal symmetry is preserved), giving possible realization of quantized Hall and spin Hall effect in three dimensions [11-15]. Upon introducing superconductivity, HOTIs can also host Majorana quasiparticles that will facilitate topological quantum computation [17]. It is crucial to note, as pointed out by refs.12-16, the gapless hinge states do not necessarily appear at all hinges of a crystal. Rather, they emerge where the two adjacent facets belong to different topological classes (*e.g.*, different gap sign). Since crystal facets are always enclosed by 1D boundaries, the gapless hinge state shall wind around a crystal and form a "loop", as illustrated in Fig. 1(c). Therefore, the existence of topological hinge state loop serves as pivotal evidence for second-order TI, but its experimental verifying is challenging as all types of hinges of a crystal need to be investigated.

Here we studied bismuth (Bi), a promising candidate of second-order 3D TI [16]. It has a rhombohedral structure [Fig. 1(d)] with space-group of $R\overline{3}m$ and $C_3$ rotation, inversion, and time-reversal symmetries. Accompanied by double band inversion at time-reversal invariant (TRI) points in $k$-space, the topological property of Bi can be described by two coupled TI phases which are gapped but with surface orientation dependent gap sign. Consequently, topological hinges states will emerge between two surfaces with opposite gap sign [16]. Experimentally, early STM studies have observed quasi-1D states on terrace edges of Bi(111) film and single crystals [18,19]. They were treated as topological edge states of a quantum spin Hall insulator [20]; while later studies show that they could be hinge states of a HOTI phase [16]. Recently, several STM studies observed more hinge states on Bi films [21-25] and nanocrystals [26-27] with different facets, some of which are considered as evidence of HOTI. However, most of these studies only observed density-of-states (DOS) peaks at the hinges, the characteristic feature of topological hinge state — dispersive 1D band with spin-helical structure, are rarely reported. Particularly, any closed loop constructed by topological hinge states has not been observed so far.

In this work, we fabricated mesoscopic Bi crystals on superconducting $V_3Si$ substrate. Using low-temperature STM, we identified various types of well-defined facets and hinges. Via quasiparticle interference (QPI) measurement, we directly observed dispersive 1D state in different hinges. To verify their topological nature, we deposited Fe clusters to induce spin-flip scattering and observed new scattering channels at certain hinges, supporting their spin-helical structure. Combining with first-principle calculation and global symmetry analysis, we show that these spin-helical hinge states formed a closed loop encircling the crystal. Our study provided the first complete verification of topological bulk-boundary correspondence and the second-order topology in Bi. We further observed proximity-induced superconducting gap in hinge states, enabling HOTI to host Majorana quasiparticles.

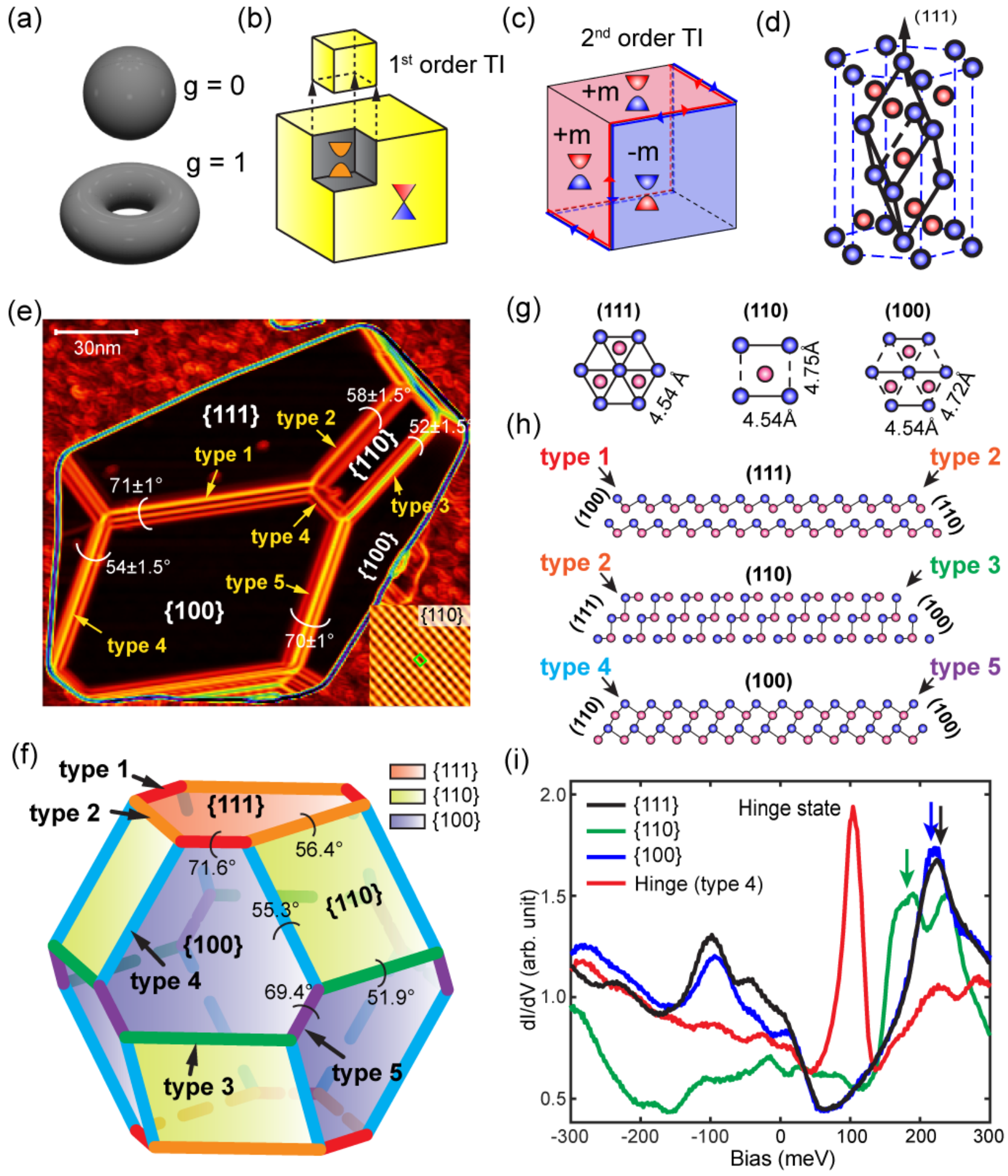

**Fig.1.** (a) Sphere and torus with different topological invariants. (b) Sketches of first-order 3D TI with topological surface state and (c) second-order TI with topological hinge states. The hinge states appear at intersections of two facets with opposite mass term, forming a closed loop. (d) Crystal structure of Bi. Blue/red spheres indicate two Bi atoms in the unit cell. (e) Topography of a Bi crystal processed by edge detection to emphasize the hinges. Different facets, hinges, and angles between adjacent facets are indicated. Inset: atomic-resolution image of (110) facet. (f) Model of Bi crystal constructed by three low index facets: {111}, {110} and {100}. Different types of facets/hinges are indicated by different colors. (g) In-plane lattice structures of three facets. Bule/red spheres indicate Bi atoms in upper/lower layer. (h) Side views of three facets and their adjacent facets. Atomic structures of all 5 types of hinges are illustrated. (i) Typical dI/dV spectra taken on different facets and type-4 hinge ($V_b = 300mV, I = 100pA$).

Mesoscopic Bi crystals were fabricated by thermal deposition of Bi on $V_3Si(111)$ surface [Part-I-1 of supplementary materials (SM)]. $V_3Si$ is a classical A15 type superconductor with a $T_c$ of 16K [28]. We have obtained clean surface of $V_3Si(111)$ [29] and utilized it as superconducting substrate. Fig. 1(e) shows topographic image of an individual Bi crystal (see Fig. S1 for additional images), which has a lateral size of 150nm×100nm and a height of 60nm. The image was processed by edge detection algorithm to visualize the hinges (Part I-3 of SM). Via examining the orientation, symmetry, and lattice constant of different facets, we determine that all Bi crystals have three kinds of low index facets: $\{111\}$, $\{110\}$ and $\{100\}$, as marked in Fig.1(e). Fig. 1(f) shows a crystal model constructed by these facets, in which we can identify five types of hinges [marked by type 1-5 in Figs. 1(e,f)]. The atomic structure of all facets/hinges are further illustrated in Figs. 1(g,h). Specifically, $\{111\}$ facet is hexagonal with a lattice constant of 0.45 nm. It is adjacent to $\{100\}$ and $\{110\}$ facets, forming type-1 and type-2 hinges. The $\{110\}$ facet has rectangular lattice with constants of 0.45 nm for short side and 0.47 nm for long side [see Fig. 1(e) inset for atomically resolved image]. The short sides are adjacent to $\{100\}$ facets, forming type-3 hinges; and one of the long sides is adjacent to $\{100\}$ facets, forming type-4 hinges. The $\{100\}$ facet has quasi-hexagonal structure. The intersection of two $\{100\}$ facets forms type-5 hinges [Fig. 1(h)]. The dihedral angles between different facets are marked in Fig.1(f), which agree with the measured angle in Fig. 1(e). We note that local structures of type-1/2 hinges are similar with edge A/B of hexagonal cavity on bulk Bi(111) surface (with the ending Bi atom pointing up and down, respectively) [19].

Fig. 1(i) shows typical dI/dV spectra acquired at different locations of Bi crystal. The spectra on $\{111\}$, $\{100\}$ and $\{110\}$ facets display DOS peaks at 220 mV, 210 mV and 185mV, respectively (indicated by arrows). These features are consistent with that observed in Bi films and single crystals with similar orientations [18-27], arising from DOS singularity of 2D surface state. In contrast, dI/dV spectra taken on crystal hinges are notably different. As represented by red curve in Fig. 1(i) (on type-4 hinge), it displays a DOS peak near 100 meV, which is absent in spectra of facets. Similar peak-like feature is also observed on other hinges in the range of ~ 90-160 meV [see left panels in Figs. 2(b-f)]. As we will show below, these peaks originate from DOS singularity of 1D hinge states. Fig. 2(a) shows several dI/dV maps of a Bi crystal with energies of 90 meV, 120 meV and 150 meV. High DOS intensities are observed only in regions near the hinges, further indicating localized 1D states; meanwhile different hinges have different contrast due to variation of peak energies. More detailed dI/dV spectra taken across the hinges are shown in Fig.S3.

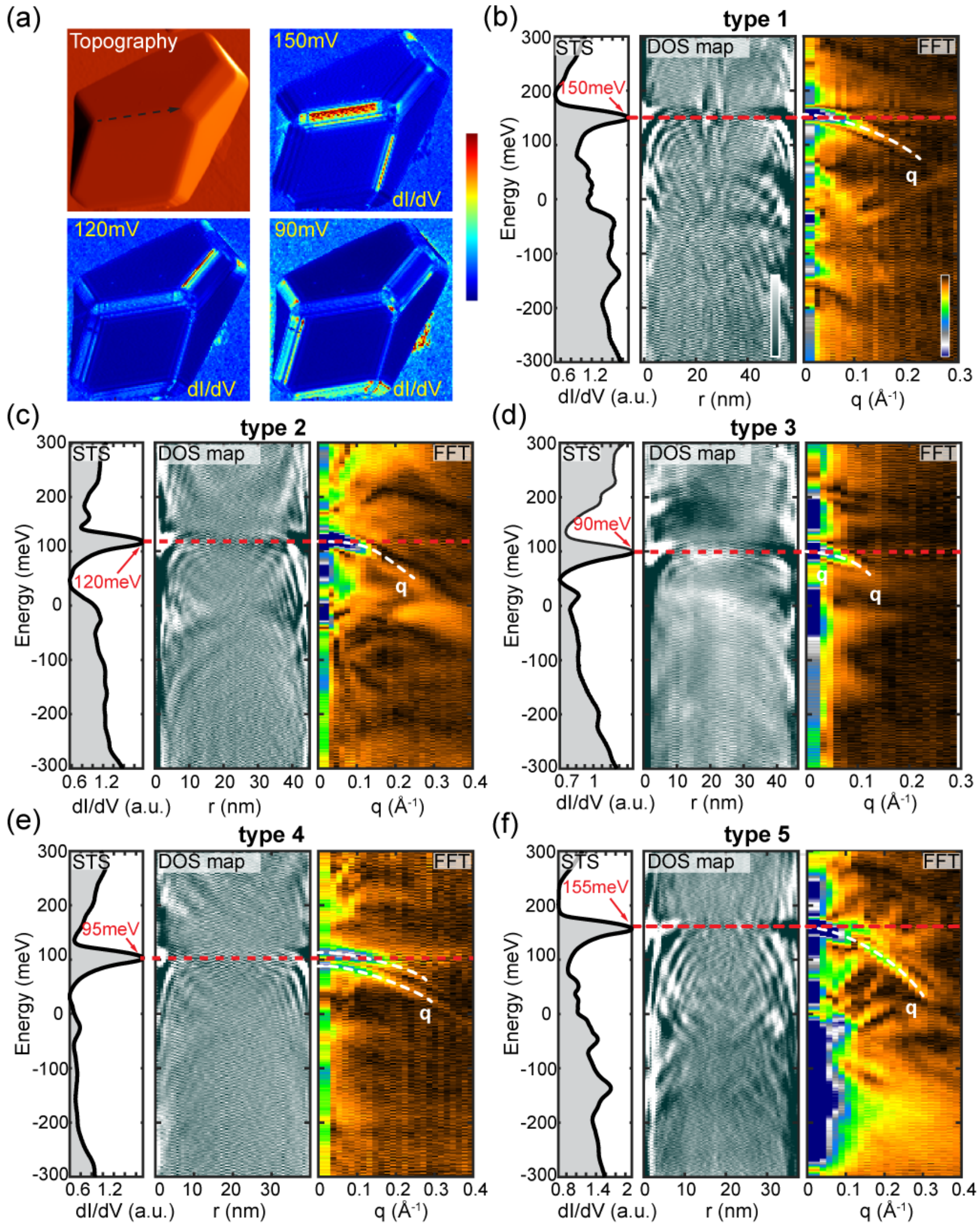

**Fig.2.** (a) Topography ($V_b = 1V, I = 10pA$) and dI/dV maps taken at energies of 1D singularity peaks. (b-f) QPI measurements on all types of hinges ($V_b = 300mV, I = 200pA$). Left panels: dI/dV spectra with DOS peaks (corresponding to band top in QPI dispersion) marked by arrows. Middle panels: Real-space dI/dV distributions (QPI patterns). Right panels: FFT of real-space distributions giving q-space dispersion of each hinge state (tracked by white dashed lines).

To investigate the dispersion of hinge states, we conducted QPI measurements by acquiring dI/dV line spectra along all the hinges (with a spatial interval ≲0.5nm). Figs. 2(b~f) summarized the QPI results of hinges types 1–5, respectively. In each figure, left panel shows representative dI/dV spectra, and middle panel shows real-space dI/dV distribution along the hinges. Clear DOS modulations are observed near the ends of hinges. Notably, the modulation periods vary with energy, indicating they arise from the interference of dispersive electrons. Right panels show the fast-Fourier transforms (FFT) of dI/dV distributions, visualizing the scattering vectors (***q***). As tracked by dashed curves, the dispersions of ***q*** are "hole-like" but have different shape and band top position for different hinges. Importantly, the band tops (***q***=0) all align with the DOS peak in dI/dV spectra. This feature is well expected for 1D-like band dispersion, which give rise to van-Hove singularity near band top. The possible contribution of projected surface states on the hinges was ruled out by comparing QPI measured on the hinges and nearby surfaces (see Part-II of SM for details).

As dispersive hinge states are observed, it is necessary to verify whether they have topological origin. For topological boundary states with spin-helical structure, nonmagnetic scatterer cannot induce backscattering between channels with opposite spin [30-33]. Fig. 3(a) illustrates the typical dispersion of spin-helical state in "type-A" edge of Bi(111) bilayer [19,20,34], which has similar structure with the type-1 hinge [Fig. 1(h)]. It has been shown that only $\boldsymbol{q}$ connects the same spin can be observed in QPI generated by nonmagnetic scatterer [19] (like the endpoints of hinges). However, magnetic impurity can locally break time-reversal symmetry (TSR) and open a gap at Dirac point [35]. This leads to the degeneration of the spin-helical structure, enabling spin-flip scattering around the magnetic impurities. As reported in ref. 24, the backscattering vector $\boldsymbol{q}^*$ in Fig. 3(a) can be observed after depositing Fe clusters on the type-A edge. Similarly, backscattering vectors were also observed on the surface of Fe doped 3D TI $Bi_2Te_3$ [36]. Therefore, magnetic scatterers can be used to distinguish topological hinge states from trivial boundary state.

We then performed QPI measurements after inducing Fe atoms as magnetic impurities (Part-I-1 of SM). A typical topographic image after Fe deposition is shown in Fig.3(b), where Fe atoms formed clusters or short chains on the hinges, generating QPI patterns around them. Figs.3(c-d) compared QPI dispersions of type-1, 4 before and after Fe deposition, respectively (results for other types of hinges are shown in Fig.S2). Notably, a new branch of scattering vectors $\boldsymbol{q}^*$ is observed in type-1 and type-4 hinges after Fe deposition (yellow dashed curves), while the original $q$ branch is still observable. $\boldsymbol{q}^*$ and $\boldsymbol{q}$ have similar band top, consistent with the dispersion shown in Fig.3(a) (lower panel). This suggests $\boldsymbol{q}^*$ originates from spin-flip scattering inducted by Fe cluster, indicating the topological origin of type-1 and 4 hinge states. In contrast, QPI dispersions of type-2,3 hinge show no significant changes after Fe deposition, suggesting they are topologically trivial. Similar results are repeatedly observed on different individual hinges of types 1-4 (see Fig. S2 for additional data). For type-5 hinges, however, due to the difficulty of finding properly positioned Fe cluster, the QPI results after Fe deposition is lacking. We will discuss type-5 hinge using symmetry analysis later.

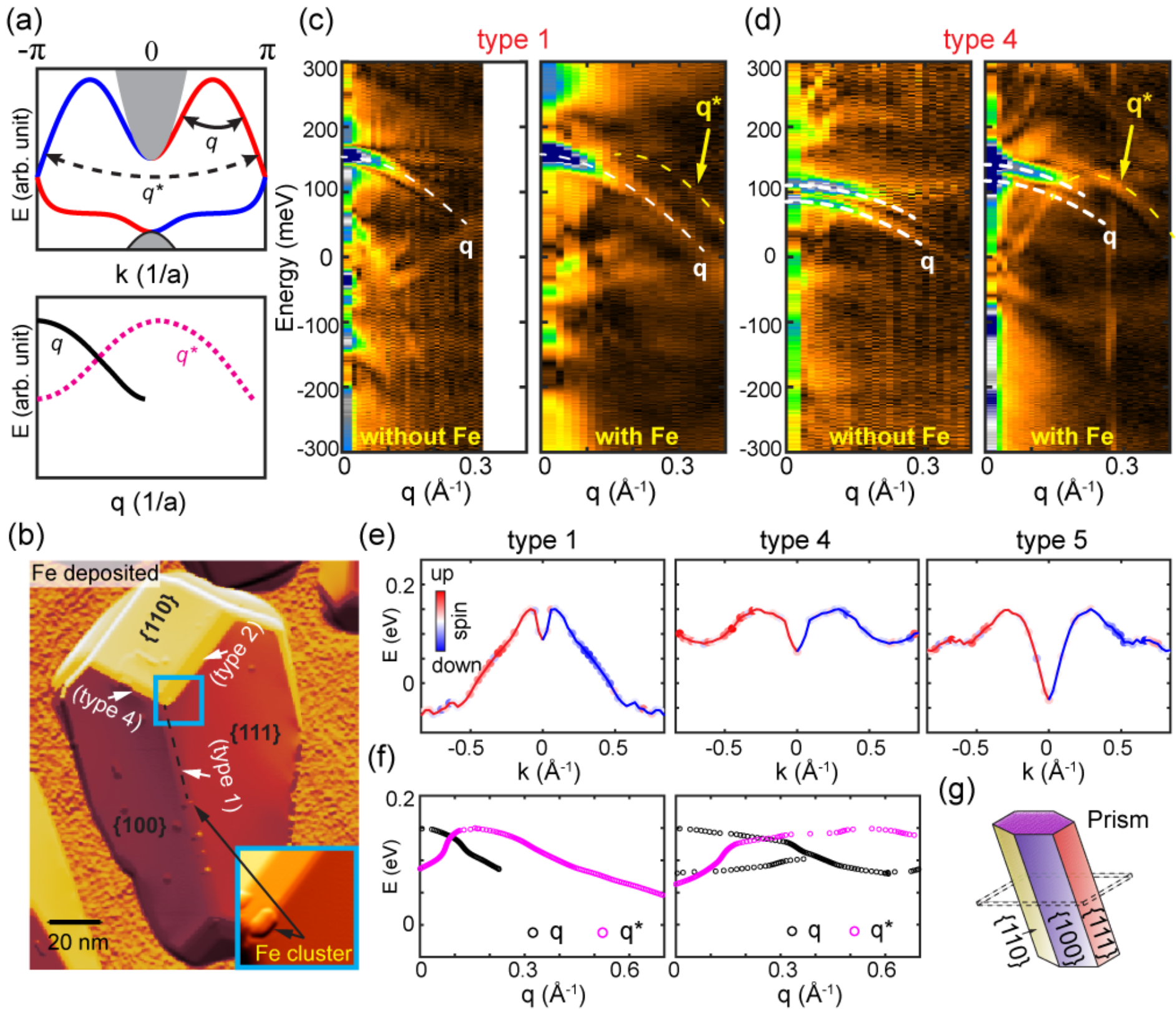

**Fig.3.** (a) Upper panel: typical dispersion of topological edge state in Bi(111) bilayer [20], red/blue represent spin-up/down. Lower panel: $\boldsymbol{q}/\boldsymbol{q}^*$ dispersion corresponding to spin-conserving/spin-flip scattering. (b) Topographic image of a Bi crystal after Fe deposition ($V_b = 1V, I = 3pA$, gradient image). Inset shows zoom-in image of a Fe cluster on hinge. (c,d) QPI dispersions of type-1,4 hinges after and before Fe deposition ($V_b = 300mV, I = 200pA$). White/yellow dashed lines indicate QPI branches of spin-conserving/spin-flip scattering $\boldsymbol{q}/\boldsymbol{q}^*$. (e) Calculated band structure and spin polarization of hinge states in type-1,4,5 hinges. Red/blue represent spin up/down and intensity at each $k$ point. (f) Calculated $\boldsymbol{q}/\boldsymbol{q}^*$ dispersions of type-1,4 hinge states. (g) Constructed prism for calculation.

To further examine the properties of the hinge states and their topological nature, we performed first-principles calculations. Normally, the DFT method is difficult to treat large open-boundary system like the mesoscopic crystals. Here we utilized a Hamiltonian Graph Neural Network model (HamGNN) [37] which can accelerate the calculation of *ab initio* Hamiltonian for large structures (see Part-III of SM). Considering type 1-3 and type 4-5 hinges are parallel to each other, respectively [Fig. 1(f)], we constructed two prisms which contain them [Fig. 3(g) and Figs. S11]. By applying HamGNN method to calculate the band structures of the prisms, we obtain dispersive 1D states with wavefunctions localized on all types of hinges. The results for type-1,4,5 hinges are shown in Figs. 3(e) and others are shown in Figs. S11. We found that all hinge states display M-shaped hole band centered at Γ, which qualitatively account for the hole-like dispersion in QPI [Figs. 2(b-f)]. Particularly, the calculated spin components of type-1,4,5 hinge states display clear polarized feature [Fig. 3(e)]; while the spin polarizations of type-2,3 hinges are significantly weaker than type-1,4,5 [Figs. S11(o)]. This suggests a spin-helical structure for type-1,4,5 hinges. The corresponding spin-conserving and spin-flip scattering vectors (***q***/***q****) of type-1,4 hinges are shown in Fig. 3(f), which qualitatively agree with QPI dispersions in Figs. 3(c-d). Therefore, our calculation supports the existence of 1D state on all the hinges but topological origin only for type-1,4,5 hinge states.

As the emergence of topological hinge state should rely on intrinsic topology of the system, we performed a symmetry analysis of Bi crystal. The topological properties of Bi (bulk and hinges) can be analyzed with classification theory of TI protected by crystalline symmetries [16]. Bi has a space group of $R\bar{3}m$ and point group of $D_{3d}$. Its bulk state belongs to symmetry class AII, classified by two topological invariants: a mirror Chern number and a $\mathbb{Z}_2$ invariant indicating HOTI, respectively [38,39]. Previous studies [16,40] of bulk Bi give 0 and 1 for these two topological invariants, indicating it is a HOTI.

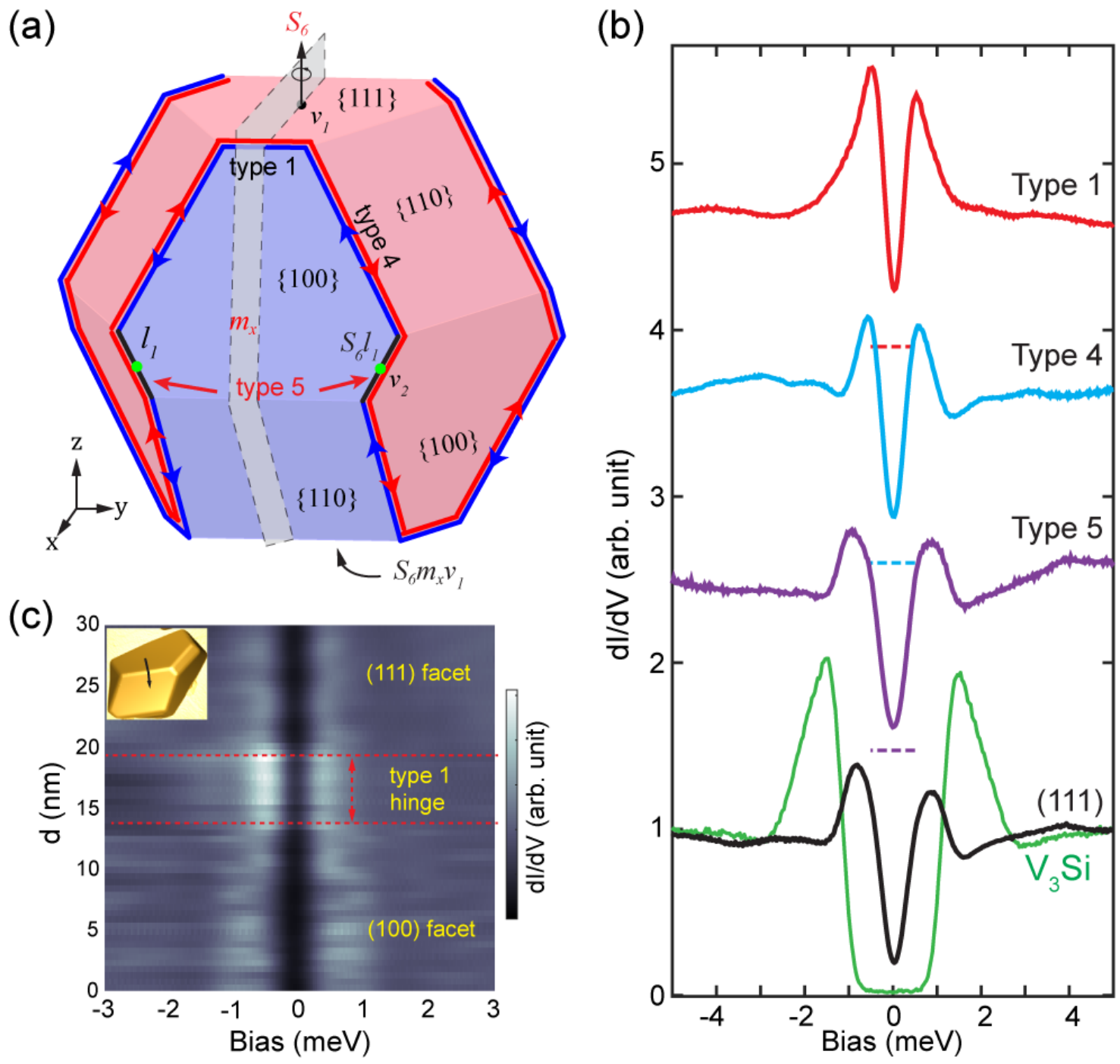

**Fig.4.** (a) Symmetry analysis of HOTI phase and topological hinge state in Bi. Symbols used in analysis are marked. The type 1,4,5 hinges with spin-helical state are marked by blue/red arrows, which form a loop and divide the surface into two separated regions with opposite mass term (indicated by red/blue colors). (b) Superconducting gap spectra on $V_3Si$ substrate, and the (111) facet and type 1,4,5 hinges of a Bi crystal ($V_b = 5mV, I = 200pA$). (c) Spatial variation of superconducting gap from (111) facet to (100) facet across type 1 hinge.

We constructed a Dirac Hamiltonian [39] from the above topological invariants and analyzed possible hinge states from its symmetry properties (see Part-IV of SM for details). The main results are summarized as following: we denote two generators of the point group $D_{3d}$ by $S_6$ (a six-fold improper rotation along axis $z$) and $m_x$ (reflection along axis $x$). The high-symmetry point $v_2$ and type-5 hinge are symmetric under $S_6$ and $m_x$, thus must be gapless. Furthermore, any path between high-symmetry points $v_1$ and $S_6 m_x v_1$ [see Fig. 4(a)] will cross an anomaly (i.e., gapless state). This means the top and bottom {111} facets must belong to two different regions separated by a loop of gapless state. And two type-5 hinges, like edges $l_1$ and $S_6 l_1$ in Fig. 4(a), must be connected by gapless hinge states. According to the above analysis and our experimental results, the only possible configuration of boundary states is shown in Fig. 4(a). The type-1,4,5 hinge states are topological and they form a loop which separates the crystal surface into two regions with opposite mass term [red/blue colored surface in Fig. 4(a)]. Spin-helical hinge states thus emerge at the intersections of these two regions. Similar argument on bulk-boundary correspondence of HOTI was also suggested in ref. 16. Therefore, our STM measurements, calculations and symmetry analysis have provided compelling evidence on the existence of higher-order topology in Bi.

Finally, as the Bi crystals are grown superconducting $V_3Si$, we observed proximity-induced superconductivity in topological hinge states. Fig. 4(b) shows typical low-energy dI/dV measured on $V_3Si$ substrate and different locations of Bi crystals. Superconducting gaps are generally observed on all facets and hinges of Bi crystal with a size Δ of ~0.4 to 0.5 meV (gap size of $V_3Si$ is ~1.3meV). The overall proximity superconductivity in Bi crystal is quite robust (Fig. S4). Fig. 4(c) shows a dI/dV linecut starts from (111) facet and ends at (100) facet, crossing a type-1 hinge. Notably, the proximity gap displays enhanced coherence peaks at the hinge region. It evidenced that the topological hinge state does have its own superconducting gap opening that is distinguishable from bulk/surface states. According to Fu-Kane model [41], a spin-helical 1D state with proximity superconductivity could host Majorana modes at its endpoints (which can be created by magnetic clusters).

In summary, we have conducted a comprehensive investigation on boundary states of Bi crystal. We have examined a complete set of hinges sufficient for constructing a 3D crystal, facilitating a global verification of higher-order bulk-boundary correspondence. The observed topological hinge states formed a closed loop and divided the crystal surface into two parts with opposite mass term, manifesting the bulk-boundary correspondence of HOTI. Therefore, our study serves as the first global verification of topological bulk-boundary correspondence and provide decisive evidence on the second-order topology in Bi. Furthermore, the Bi crystal on $V_3Si$ provided a unique HOTI/superconductor hybrid system. The proximity superconductivity induced in topological hinge states makes it a potential platform to realize Majorana quasiparticles.

**References:**

[1]. Nakahara, M., *Geometry, Topology and Physics* (Hilger, Bristol, 1990).
[2]. Thouless, D. J., M. Kohmoto, M. P. Nightingale, and M. den Nijs, *Quantized Hall Conductance in a Two-Dimensional Periodic Potential*, Phys. Rev. Lett. 49, 405 (1982).
[3]. Wen, X. G., *Topological orders and edge excitations in fractional quantum Hall states*, Adv. Phys. 44, 405 (1995).
[4]. Kane, C. L., and E. J. Mele, *Z2 Topological Order and the Quantum Spin Hall Effect*, Phys. Rev. Lett. 95, 146802 (2005).
[5]. Bernevig, B. A., and S. C. Zhang, *Quantum Spin Hall Effect*, Phys. Rev. Lett. 96, 106802 (2006).

[6]. Fu, L., C. L. Kane, and E. J. Mele, *Topological Insulators in Three Dimensions*, Phys. Rev. Lett. 98, 106803 (2007).
[7]. Hasan, M. Z. and C. L. Kane, *Colloquium: Topological insulators*, Reviews of Modern Physics 82, 3045 (2010).
[8]. Qi, X. L. and S. C. Zhang, *Topological insulators and superconductors*, Reviews of Modern Physics 83, 1057 (2011).
[9]. Zhang, H., et al., *Topological insulators in $Bi_2Se_3$, $Bi_2Te_3$ and $Sb_2Te_3$ with a single Dirac cone on the surface*, Nature Physics 5, 438 (2009).
[10]. Fu, L., *Topological crystalline insulators*, Phys. Rev. Lett. 106, 106802 (2011).
[11]. Benalcazar, W. A., et al., *Quantized electric multipole insulators*, Science 357, 61 (2017).
[12]. Song, Z., et al., *(d-2)-Dimensional Edge States of Rotation Symmetry Protected Topological States*, Phys. Rev. Lett. 119, 246402 (2017).
[13]. Langbehn, J., et al., *Reflection-Symmetric Second-Order Topological Insulators and Superconductors*, Phys. Rev. Lett. 119, 246401 (2017).
[14]. Schindler, F., et al., *Higher-order topological insulators*, Science Advances 4, aat0346 (2018).
[15]. M. Geier, L. Trifunovic, M. Hoskam, P. W. Brouwer, *Second-order topological insulators and superconductors with an order-two crystalline symmetry*, Phys. Rev. B 97, 205135 (2018)
[16]. Schindler, F., et al., *Higher-Order Topology in Bismuth*, Nat. Phys. 14, 918 (2018).
[17]. A. Y. Kitaev, *Fault-tolerant quantum computation by anyons*, Ann. Phys. 303, 2–30 (2003).
[18]. F. Yang, et al., *Spatial and Energy Distribution of Topological Edge States in Single Bi(111) Bilayer*, Phys. Rev. Lett. 109, 016801 (2012).
[19]. Drozdov, I. K., et al., *One-dimensional topological edge states of bismuth bilayers*, Nature Physics 10, 664 (2014).
[20]. M. Wada, S. Murakami, F. Freimuth, and G. Bihlmayer, *Localized edge states in two-dimensional topological insulators: Ultrathin Bi films*, Phys. Rev. B 83, 121310(R) (2011).
[21]. L. Peng, J.-J. Xian, P. Tang, A. Rubio, S.-C. Zhang, W. Zhang, and Y.-S. Fu, *Visualizing topological edge states of single and double bilayer Bi supported on multibilayer Bi(111) films*, Phys. Rev. B 98, 245108 (2018).
[22]. Nayak, A. K., et al., *Resolving the topological classification of bismuth with topological defects*, Science Advances 5, eaax6996 (2019).
[23]. Jack, B., et al., *Observation of a Majorana zero mode in a topologically protected edge channel*, Science 364, 1255 (2019).
[24]. Jack, B., et al., *Observation of backscattering induced by magnetism in a topological edge state*, Proc Natl Acad. Sci U S A 117, 16214 (2020).
[25]. Aggarwal, L., et al., *Evidence for higher order topology in Bi and $Bi_{0.92}Sb_{0.08}$*, Nat. Communi. 12, 4420 (2021).
[26]. Yang, F., et al., *Edge states in mesoscopic Bi islands on superconducting Nb(110)*, Phys. Rev. B 96, 235413 (2017).
[27]. Zhang, T.Z, et al., *STM observation of the hinge-states of bismuth nanocrystals*, Phys. Rev. B 108, 085422 (2023).
[28]. Hardy G.F., Hulm J.K., *Superconducting silicides and germanides*, Phys. Rev. 89, 884 (1953).
[29]. Ding, S., et al., *Surface structure and multigap superconductivity of $V_3Si$ (111) revealed by scanning tunneling microscopy*, Quantum Frontiers 2, 3 (2023).

[30]. Roushan, P., et al., *Topological surface states protected from backscattering by chiral spin texture*. Nature 460, 1106 (2009).

[31]. Zhang, T., et al., *Experimental demonstration of topological surface states protected by time-reversal symmetry*, Phys. Rev. Lett. 103, 266803 (2009).

[32]. Alpichshev, Z., et al., *STM imaging of electronic waves on the surface of $Bi_2Te_3$: topologically protected surface states and hexagonal warping effects*, Phys. Rev. Lett. 104, 016401 (2010).

[33]. Beidenkopf, H., et al., *Spatial fluctuations of helical Dirac fermions on the surface of topological insulators*, Nature Physics 7, 939 (2011).

[34]. Hofmann, P., *The surfaces of bismuth: Structural and electronic properties*, Progress in Surface Science 81, 191 (2006).

[35]. Q. Liu, C.X. Liu, C. Xu, X.Liang Qi, and S.C. Zhang, *Magnetic Impurities on the Surface of a Topological Insulator*, Phys. Rev. Lett. 102, 156603 (2009)

[36]. Y. Okada, et al., *Direct Observation of Broken Time-Reversal Symmetry on the Surface of a Magnetically Doped Topological Insulator*, Phys. Rev. Lett. 106, 206805 (2011)

[37]. Zhong, Y., Yu, H., Su, M., Gong, X. & Xiang, H., *Transferable equivariant graph neural networks for the Hamiltonians of molecules and solids*, npj Comput. Mater. 9, 182 (2023).

[38]. Eyal Cornfeld and Adam Chapman, *Classification of crystalline topological insulators and superconductors with point group symmetries*, Phys. Rev. B, 99, 075105 (2019).

[39]. Zhida Song, Sheng-Jie Huang, Yang Qi, Chen Fang, and Michael Hermele, *Topological states from topological crystals*, Science Advances, 5, eaax2007 (2019).

[40]. Hsu, C. H., et al., *Topology on a new facet of bismuth*, Proc. Natl. Acad. Sci. USA 116, 13255 (2019).

[41]. Liang Fu and C. L. Kane, *Superconducting Proximity Effect and Majorana Fermions at the Surface of a Topological Insulator*, Phys. Rev. Lett. 100. 096407 (2008).

**Acknowledgments:**

We thank professors Zhenyu Zhang and Jing Wang for helpful discussion. This work is supported by Innovation Program for Quantum Science and Technology (Grant No. 2021ZD0302803), National Natural Science Foundation of China (Grants Nos.: 92065202, 12225403, 12188101, 92365302, 12304181, 12104094, 12374144), The New Cornerstone Science Foundation (Grant No. NCI202211), Shanghai Municipal Science and Technology Major Project (Grant No. 2019SHZDZX01), Shanghai Science and Technology Program (No. 23JC1400900), Guangdong Major Project of the Basic and Applied Basic Research (Future functional materials under extreme conditions--2021B0301030005), Shanghai Pilot Program for Basic Research (Fudan University, Grant No. 21TQ1400100).

# Supplementary Material for
# "Revealing Higher-Order Topological Bulk-boundary Correspondence in Bismuth Crystal with Spin-helical Hinge State Loop and Proximity Superconductivity"

Dongming Zhao†, Yang Zhong†, Tian Yuan†, Haitao Wang, Tianxing Jiang, Yang Qi*, Hongjun Xiang, Xingao Gong*, Donglai Feng*, Tong Zhang*

† These authors contributed equally.

## Part. I: Methods and additional STM data.

### I-1. Sample preparation and STM measurement:

The sample growth and STM experiment were conducted in a low-temperature STM system (UNISOKU 1300) equipped with an MBE chamber under a vacuum of ~$1\times10^{-10}$ mbar. The $V_3Si$ (111) single crystal (Mateck GmbH) substrate was treated by repeated Argon sputtering and post-annealing at 800°C. Bi was deposited on $V_3Si$ (111) with a nominal thickness of 6 nm (calculated from the flux rate). The substrate was kept at room temperature during the deposition, followed by post-annealing at 350°C for 40min and cooling down with an average rate of 7min/°C to acquire mesoscopic Bi crystal with well-defined facets. Fe clusters were formed by depositing 0.005 monolayer Fe on Bi/$V_3Si$ at 300°C. All STM measurements were done at 4.3K except the superconducting gap measurements, which were done at 0.4K.

### I-2. Additional STM data:

Fig. S1(a) shows large-scale topographic image of mesoscopic Bi crystals grown on $V_3Si$ (111) substrate. Fig. S1(b-j) show the images of individual Bi crystals before and after Fe deposition.

Fig. S2 shows the comparison of QPI data set (including dI/dV spectra, real-space DOS map, and FFT images) taken at different hinge of types 1-4, before and after Fe deposition. It's seen that the same type of hinges always display similar dispersion of $q$ (regardless of Fe deposition). While after Fe deposition, different type-1and type-4 hinges all display similar $q^*$, respectively.

Fig. S3 shows the dI/dV spectra across different types of hinges.

Fig. S4 shows the Spatial variation of superconducting gaps from substrate to a (111) facet.

### I-3. The "edge detection" processing of STM image:

The "edge detection" algorithm of an image used a "local non-linearity" function to visualize areas that are locally non-planar (*e.g.*, the edges). Specifically, it fits a plane through a circular neighborhood at each pixel of STM image (with a radius of 2.5 pixels):

$$z_{\text{fit}} = b_x x + b_y y + \text{constant}$$

$x$ and $y$ are the coordinates of each pixel, $b_x$ and $b_y$ are the plane coefficients. Then the "local non-linearity" is expressed as:

$$L = \sqrt{\sum_n \frac{\Delta z(n)^2}{1 + {b_x}^2 + {b_y}^2}}$$

Here $n$ goes through all neighboring pixels and $\Delta z(n) = z_{\text{data}}(n) - z_{\text{fit}}(n)$ is the difference between the STM height data and the fitted plane. This formular gives large value near the edges and helps to visualize the crystal hinges. An example of STM image processed by this algorithm is shown in Fig. S5.

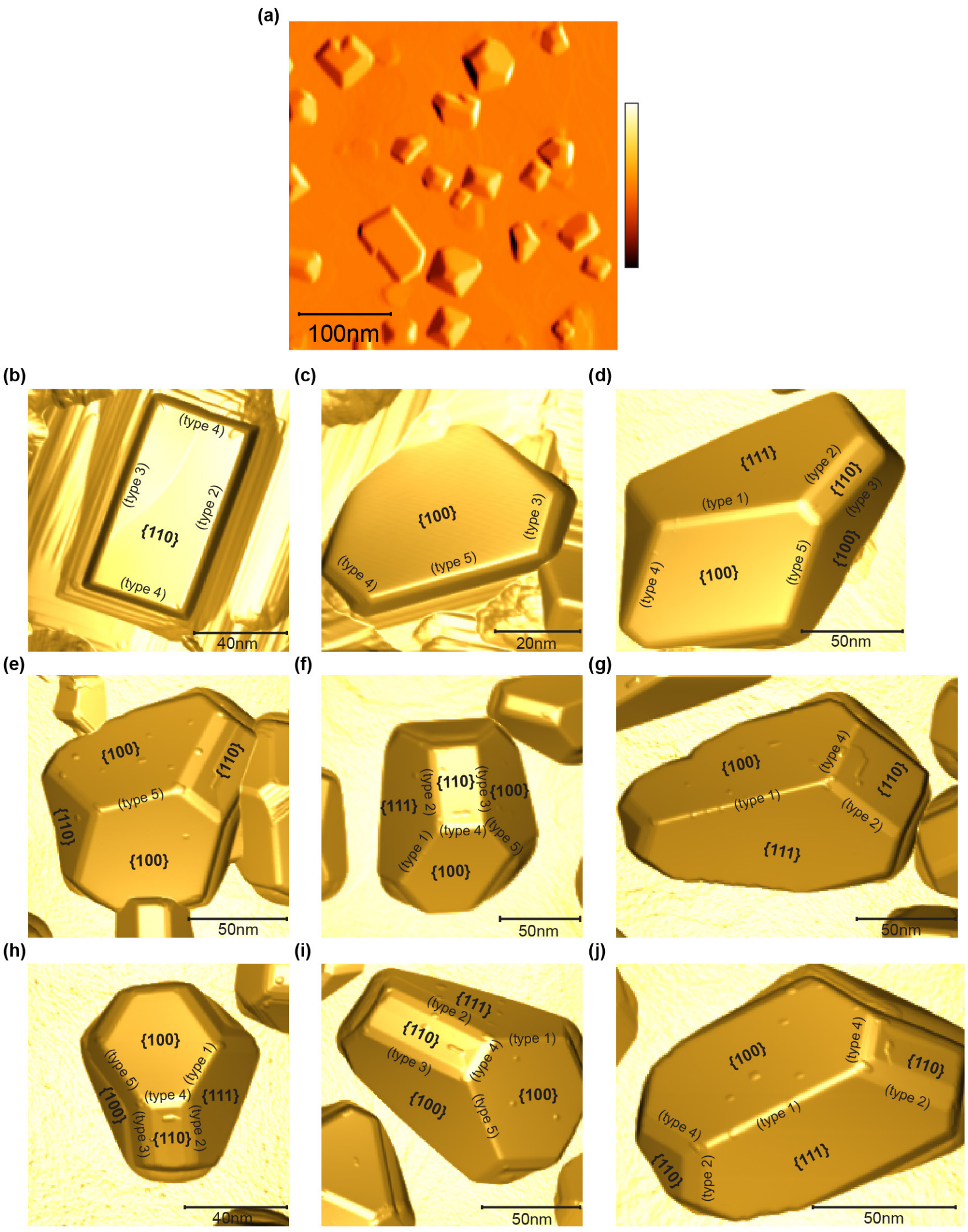

**FIG. S1**. **Additional images of Bi crystals (differential images)** (a) Large-scale topographic image of mesoscopic Bi crystals grown on V3Si(111) ($V_b = 1V, I = 10pA$). (b-d) Bi crystals without Fe depositions. (e-j) Bi crystals with Fe depositions. The indexes of different facets are labeled in the image.

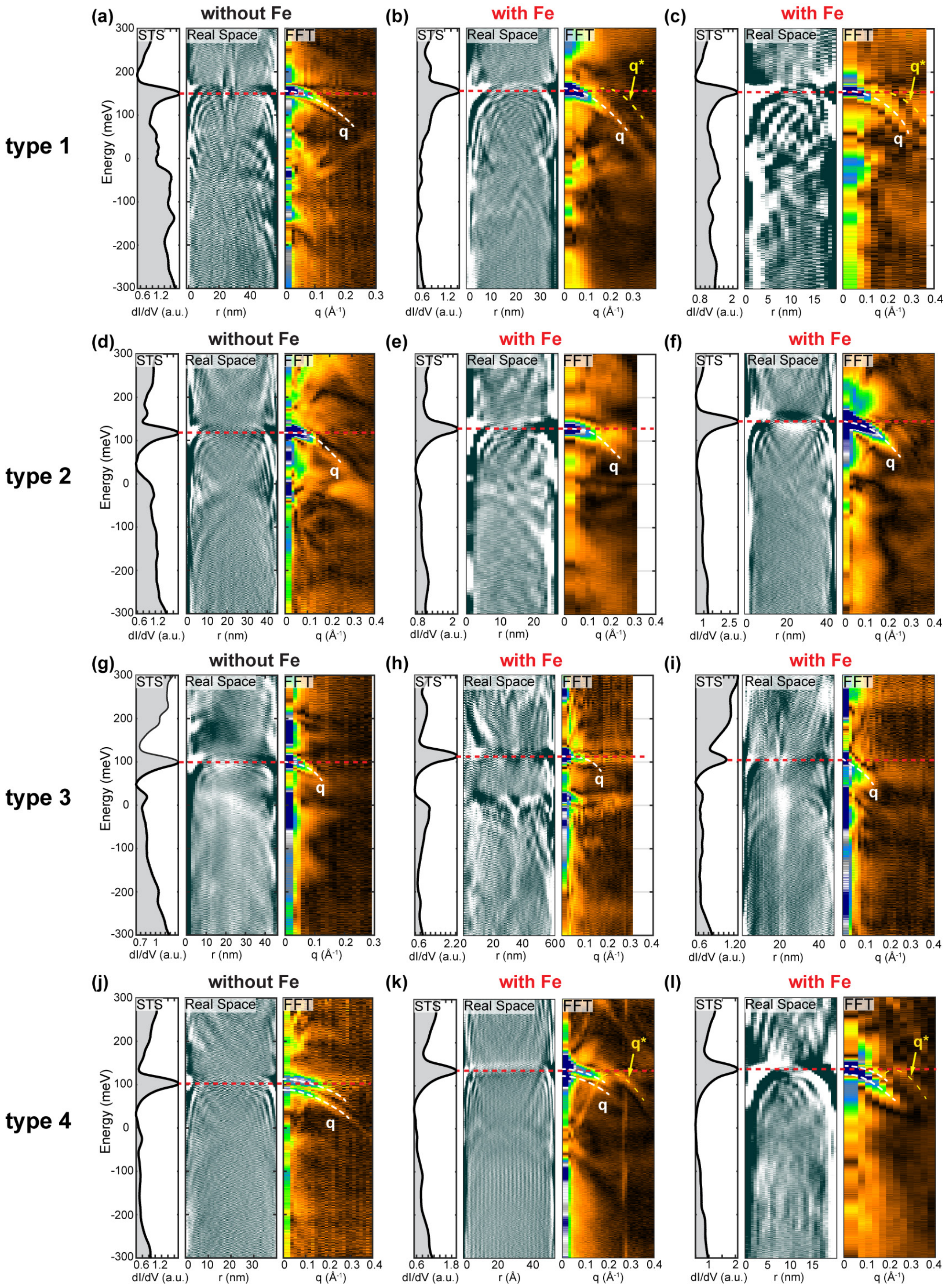

**FIG.S2. QPI Measurements on different hinges of types 1-4, before and after Fe cluster deposition (labeled above each panel).** It's seen that the same type of hinges always display similar dispersion of $q$ (regardless of Fe deposition). While after Fe deposition, different type-1and type-4 hinges all display similar $q*$, respectively.

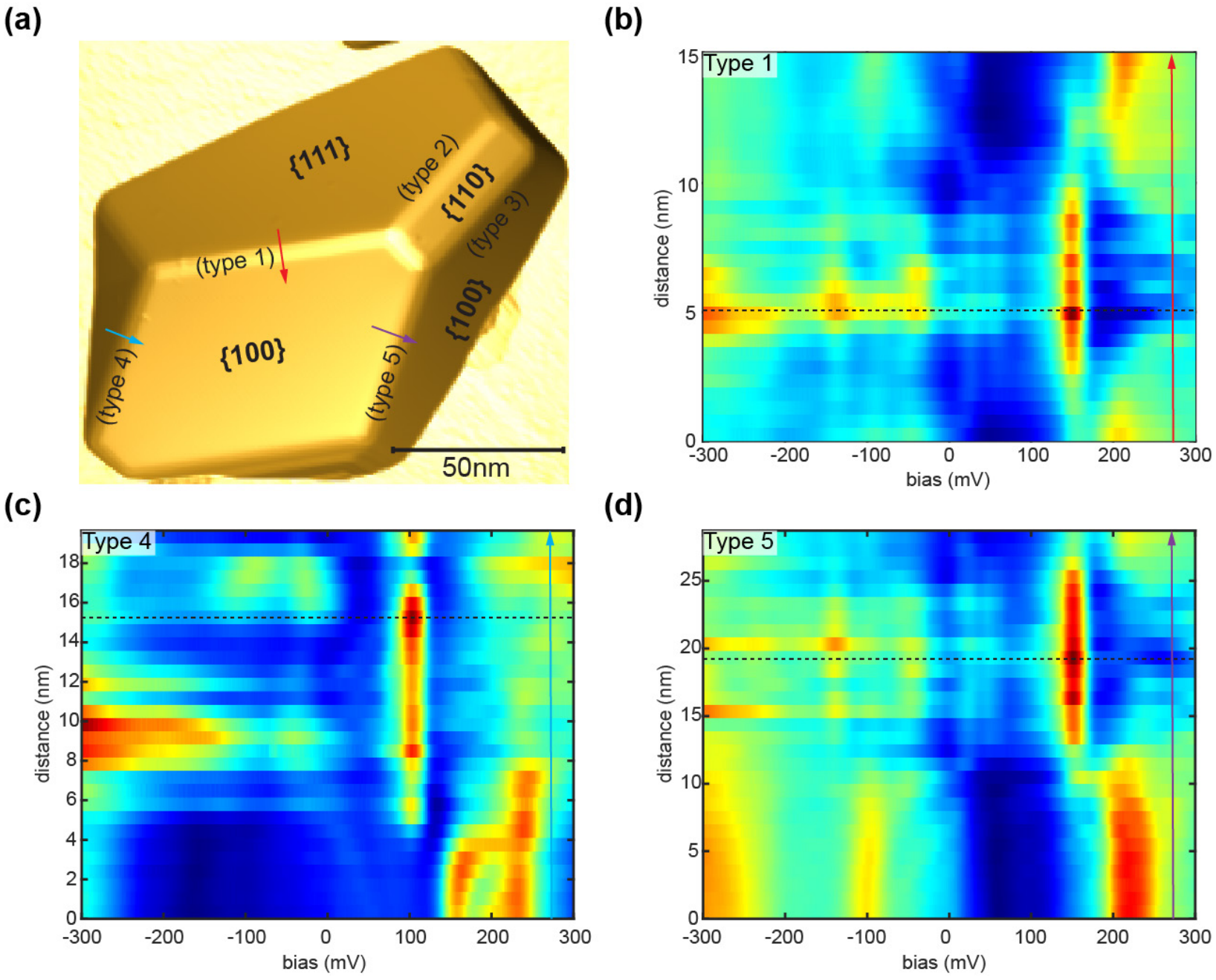

**FIG. S3. dI/dV spectra across different types of hinges.** (a) Topographic image of Bi crystal ($V_b = 1V, I = 10\ pA$). (b-d) dI/dV linecut spectra taken across type-1 (b), type-4 (c) and type-5 (d) hinges, along the arrows in (a). Black dashed lines marked the positions of the hinges. The 1D singularity DOS peaks are well localized on hinges with a full width of 5-10 nm. (Setpoint: $V_b = 300mV, I = 200pA$).

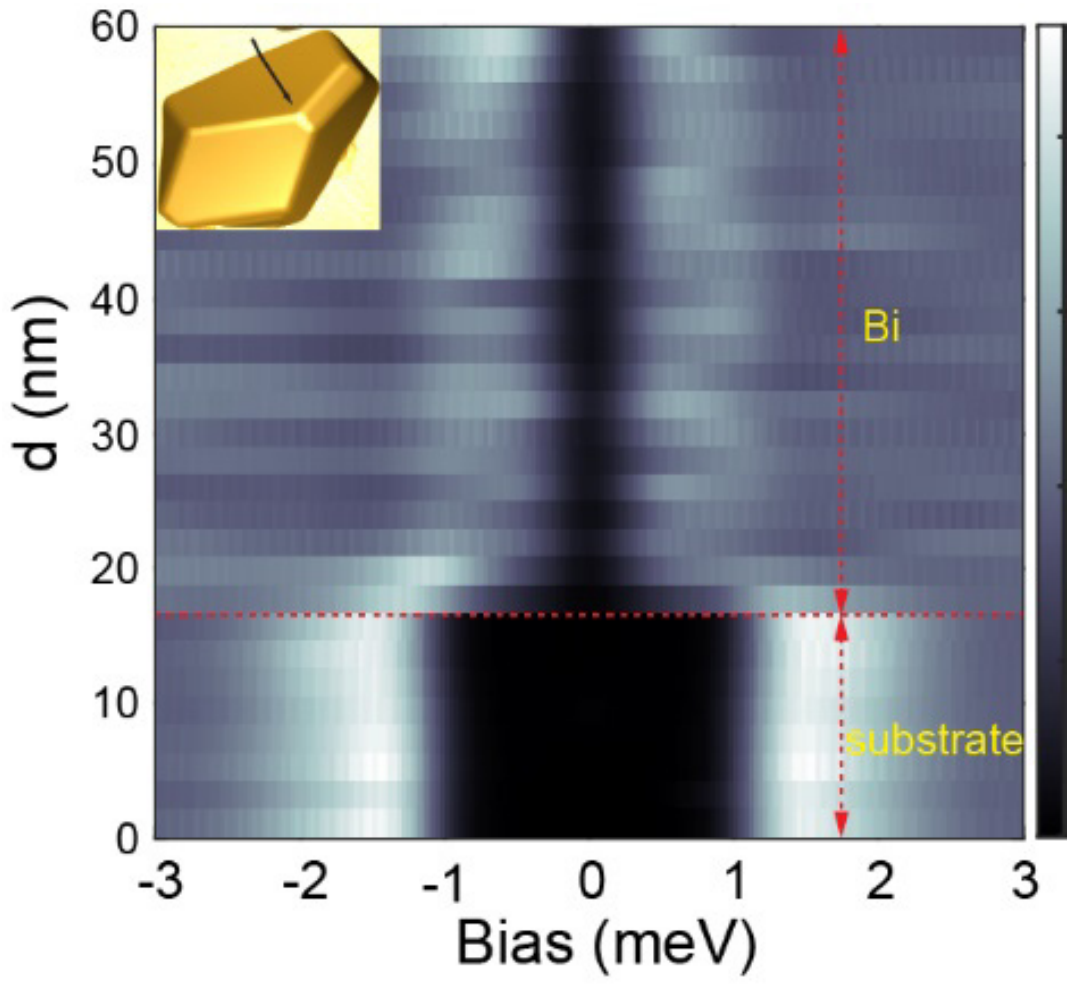

**FIG. S4.** Spatial variation of superconducting gaps from $V_3Si$ substrate to a Bi (111) facet. ($V_b = 3mV, I = 200pA$).

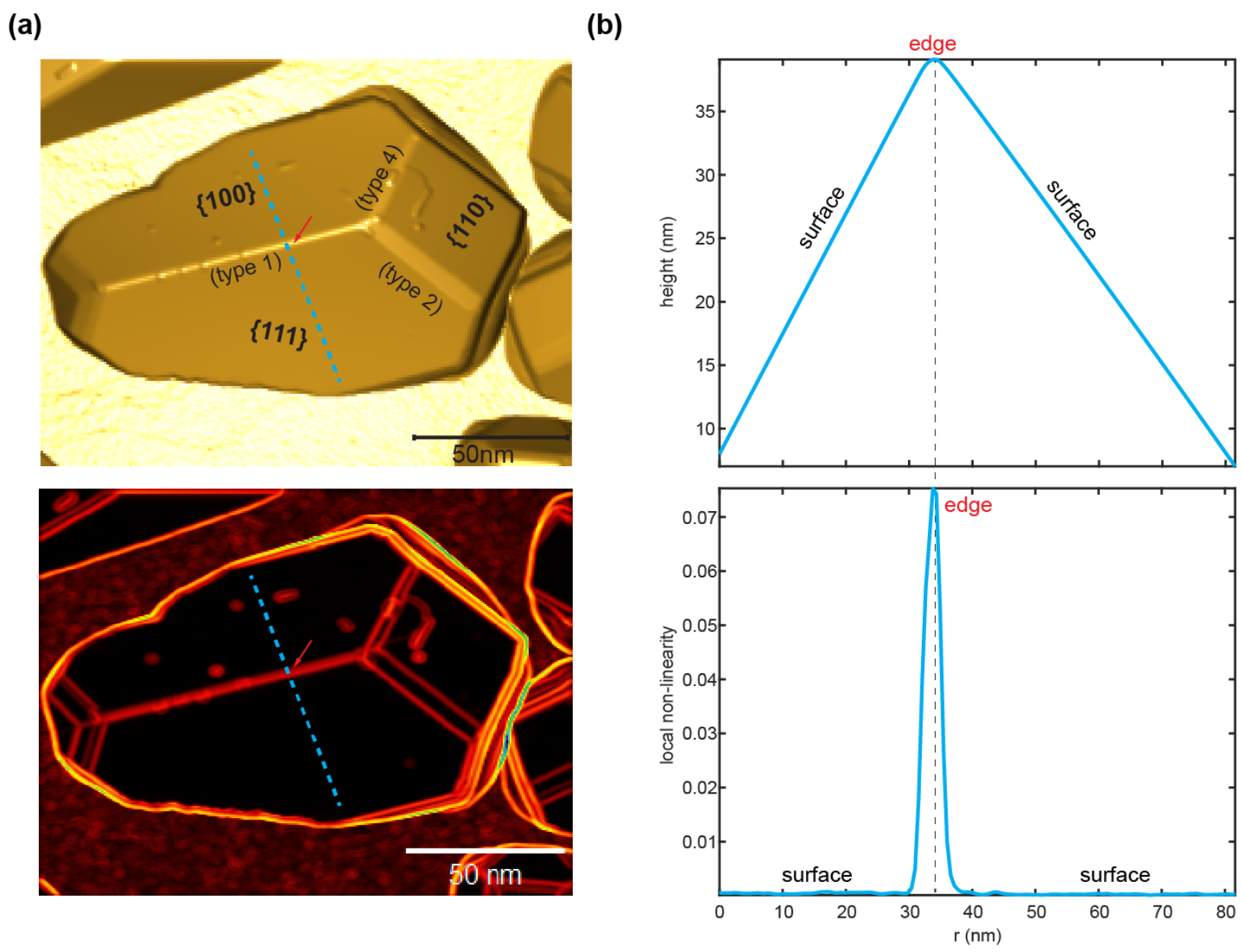

**Fig. S5**. **The edge detection processing of STM image** (a) Topographic image (up panel) and "edge detection" processed image (down panel) of a Bi crystal. (b) Extracted height (up panel) and "local non-linearity" data (down panel) from dashed blue line in (a). The position of edges in height data correspond to the peak in "local non-linearity" data.

## Part. II: Examining the surface state projection on hinges.

As Bi crystal should have surface states on different facets, it is important to investigate possible surface state projection on the hinges and distinguish them from hinge state. To address this, we measured dI/dV linecut spectra on the adjacent surfaces of type 1, 4 hinges, along trajectories parallel to the hinge. As illustrated in Fig. S6(a), if the surface states from (111) or (100) facets have projections on type-1 hinge (red arrow), they should also project onto cut 1 or cut 2 (green and blue arrows). However, the 1D QPI pattern acquired on cut 1 and 2 [Fig. S6(d,e)] are clearly different from those on type-1 hinge (either with Fe deposited or not) [Fig. S6(b,c)]. The dispersions of surface states are distinct to $q$ and $q*$ observed on type-1 hinge. Similar behaviors are also seen for the type-4 hinge, as shown in Fig. S7. These results indicate the QPI on type-1,4 hinges are dominated by 1D hinge states, which effectively rule out significant contributions from surface state QPI.

Here we did not observe the $q2$ feature reported in Ref. 19 (Nat. Phys. 10, 664(2014)). This discrepancy could be due to that type-1 hinge is the intersection of two facets, where two different surface states could encounter and hybridize; while the type-A edge in Ref. 19 is a terrace edge on (111) surface. The surface-state features measured on them are expected to be different.

We also exclude the possibility that after Fe deposition, the surface-state QPI generated by Fe atoms may affect the hinge states. In Fig. S8 below we show additional images of Bi crystals after Fe deposition (with edge enhanced images). The Fe clusters are marked by circles and arrows. Dashed lines show the locations where the QPI "with Fe" were measured. It is evident that except Fe clusters positioned directly on the hinges, which act as magnetic scatterers, other

Fe clusters on adjacent facet are typically more than 10 nm away from the hinges. At such distances, surface Fe clusters cannot directly affect the hinge states which are strongly localized on hinges with a width of ~5 nm (see Fig. 2a and Fig. S3). Even if these Fe clusters generated surface states QPI, there is no evidence of their projections on the hinge. This is because the observed QPI patterns on hinges always surround the ends or clusters on the hinge, as seen in all the real-space DOS map in Figs. S2, and more clearly demonstrated in Fig. S8(g-l) (see figure caption). If there were surface state projections from nearby Fe clusters, we would expect to detect them in the middle region of the linecut; however, no such evidence has been observed.

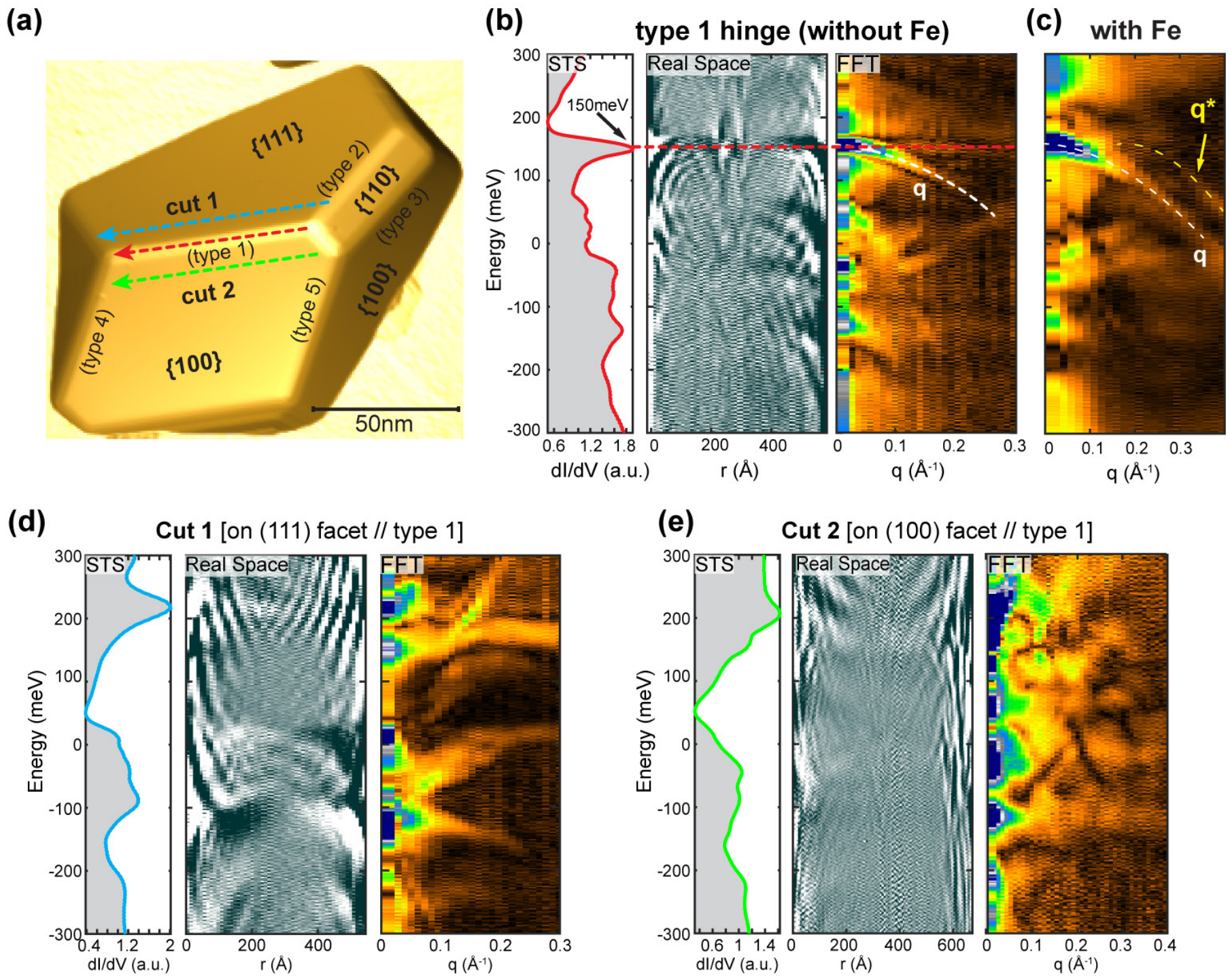

**Fig. S6. Comparison of 1D QPI taken on type 1 hinge and its adjacent surfaces.** (a) Topographic image of a Bi crystal. (b) The QPI data set taken on type-1 hinge (red arrow) in panel (a). (c) QPI data taken on a type-1 hinge with Fe deposited. (d,e) QPI data sets taken along cut 1,2 in the adjacent (111) and (100) surface, respectively, which are in parallel with type-1 hinge [bule and green arrows in panel (a)].

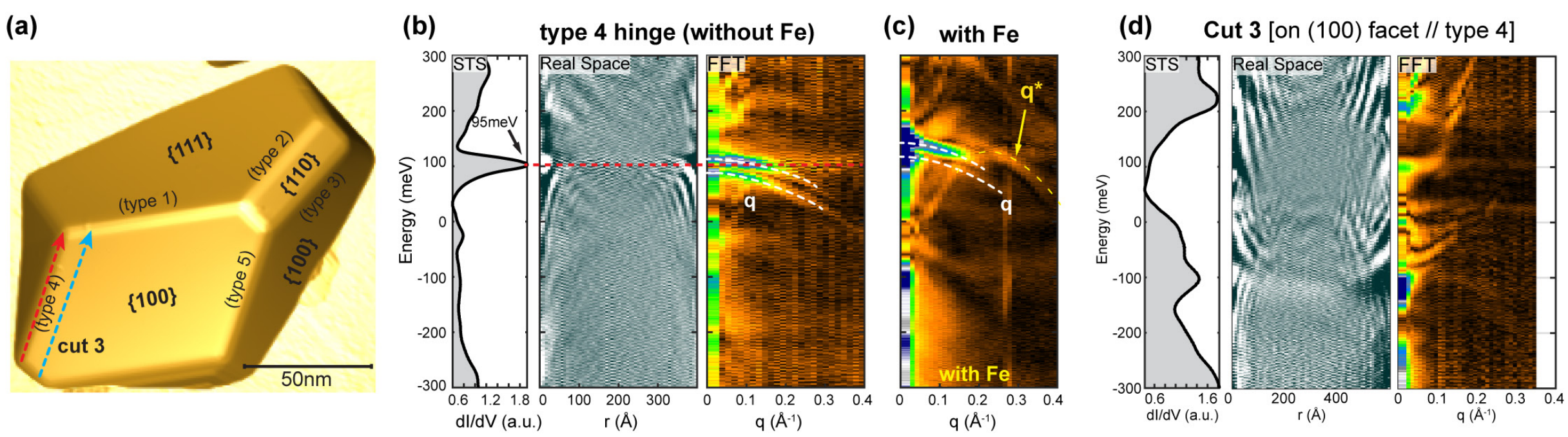

**Fig. S7. Comparison of 1D QPI taken on type 4 hinge and its adjacent surface.** (a) Topographic image of a Bi crystal. (b) The QPI data set taken on type 4 hinge in panel (a) (without Fe deposition). (c) The QPI dispersion taken on a type 4 hinge with Fe deposition. (d) The QPI data set taken along the blue arrow in panel (a).

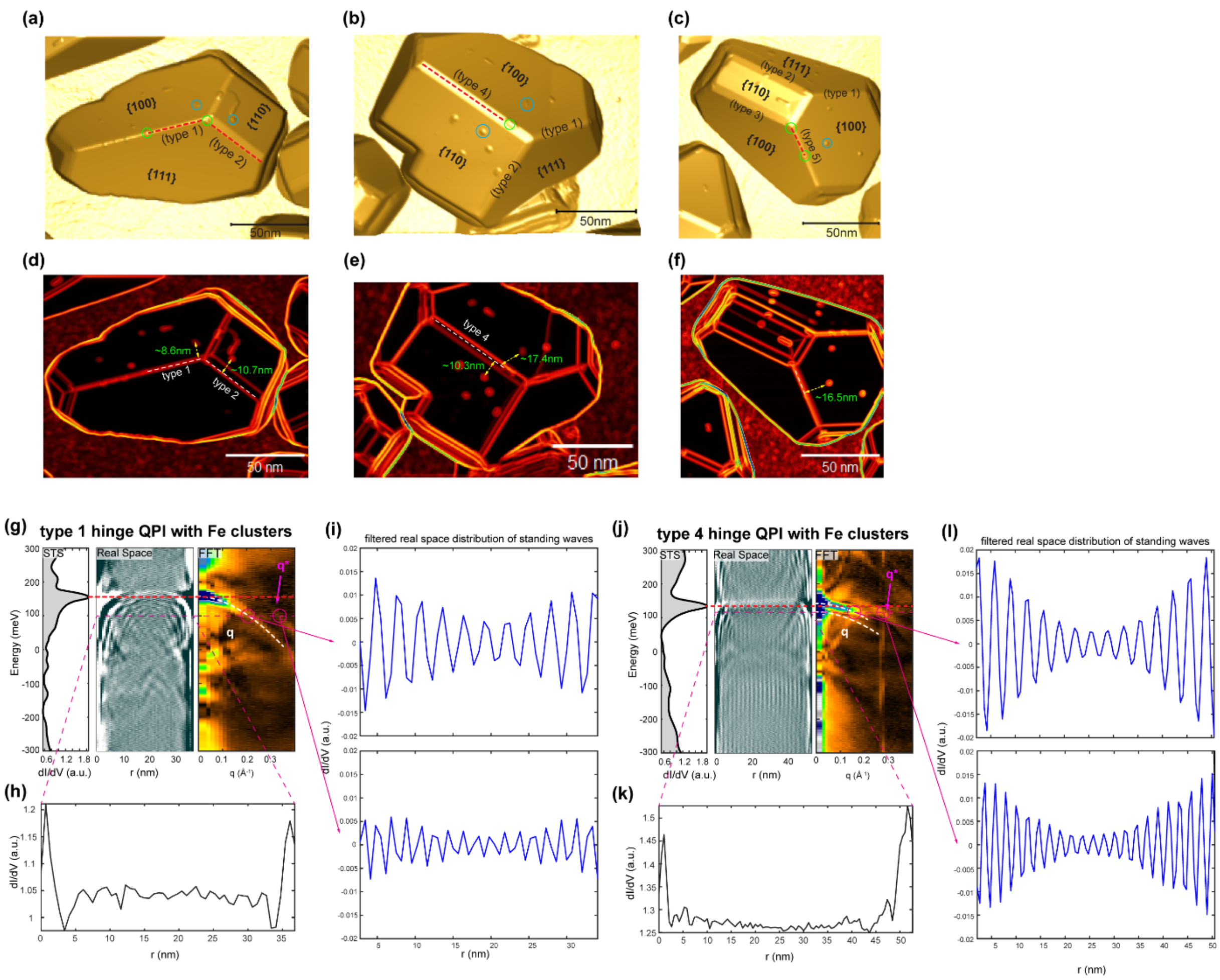

**Fig. S8.** (a-c) Topographic images of Bi crystals with Fe cluster deposited. Red dashed lines are where the QPI were measured. The Fe clusters on the end of these lines are marked by green circles, and the nearby Fe clusters on the surface are marked by blue circles. (d-f) Edge enhanced image of panel (a-c), respectively. The Fe clusters and hinges are more clearly seen. The distances between nearest clusters and the hinges are labeled. (g, j) STS and QPI measurements on type 1, 4 hinges with Fe clusters. (h, k) Raw data of dI/dV distribution along the hinge, at E = 100meV. The QPI modulation is more pronounced at the end of hinges or Fe cluster. (i, l) Filtered dI/dV distribution with q and q*, as marked by pink circles in (g) and (j). Upper/Lower panels correspond to q/q* wavevectors. The modulations of q/q* are also pronounced at the end of hinges or Fe cluster, indicating they are from the scattering of hinge states.

## Part. III: Band structure calculation of Bi hinge states:

To investigate topological boundary states with open boundary conditions, it is necessary to construct sufficiently large nanoscale systems with vacuum interfaces, such as 2-dimensional slabs and 1-dimensional bars. Directly employing Density Functional Theory (DFT) methods for electronic structure calculations of these large open systems with thousands of atoms is impractical. Currently, the conventional approach involves constructing open systems that possess a certain high point group symmetry and approximate their electronic structures around a high-symmetry *k* point in reciprocal space using effective parameterized Hamiltonians [1-3].

For instance, Schindler et al. [4] derived a Dirac model with $C_3$ and inversion symmetries for the Higher-Order Topological Insulator (HOTI) states of Bismuth, starting from the (BHZ) model [5] for three-dimensional topological insulators. However, this method relies on intricate symmetry analyses and proves challenging to extend to general HOTI states. When the boundary states of a system reside on interfaces with lower symmetry, it is challenging to construct a reasonable parameterized Hamiltonian solely based on symmetry considerations.

Recently, we proposed a Hamiltonian graph neural network model (HamGNN) [6] that can directly map crystal structures to *ab initio* tight-binding Hamiltonian matrices, bypassing the expensive self-consistent iteration process and greatly accelerating the calculation of *ab initio* Hamiltonian matrices for large structures. HamGNN eschews any reliance on empirical assumption. Instead, it constructs authentic Hamiltonians by directly learning the interactions of atoms in their local environments through the powerful fitting ability of deep neural networks. Consequently, HamGNN is well-suited for investigating various edge states. In this work, we utilized HamGNN to study the electronic structure of the hinge states of mesoscopic Bi crystal.

We constructed 200 one-dimensional prisms of Bi extending along five directions to train the HamGNN model. Each prism contains no more than one hundred atoms. The tight-binding Hamiltonian matrices of these prisms were calculated using OpenMX [7,8] and were divided into training, validation, and test sets in a ratio of 0.8:0.1:0.1 respectively. Fig. S9 shows the comparison between the Hamiltonian matrix elements calculated by the trained HamGNN model and OpenMX for the Bi prisms in the test set. The mean absolute error (MAE) of the predicted Hamiltonian matrix elements by HamGNN on the testing set is merely 0.96 meV, demonstrating its high accuracy and reliability in reproducing electronic structures. Furthermore, it is worth noting that HamGNN achieves an MAE value of only 8.3 meV for energy bands near the Fermi levels. Fig. S10 shows the energy bands of a Bi prism calculated by HamGNN and OpenMX for a test, the calculated dispersions well match with each other.

Utilizing this trained HamGNN model, Figs. S11(a,d) show the calculated full band structure of the two prisms extended along directions parallel to type 1,2,3 and type 4,5, respectively. The bands with wavefunction localized on the hinges are recognized as the hinge state [Figs. S11(b,c,e,f)]. It is worth noting that hinge states of type 2 and 3 share the same localized wavefunction as shown in Fig. S11(c). The hinge state dispersion and projected spin component (along its maximum direction) are shown in Figs. S11(g-j) and (k-n). The averaged spin polarization, calculated by averaging ($S_k$ - $S_{-k}$) over k, is shown in Fig. S11(o). It is seen that the polarization of type 1,4,5 hinges are significantly large than that of type 2,3 hinges, suggesting a spin-helical structure of the former.

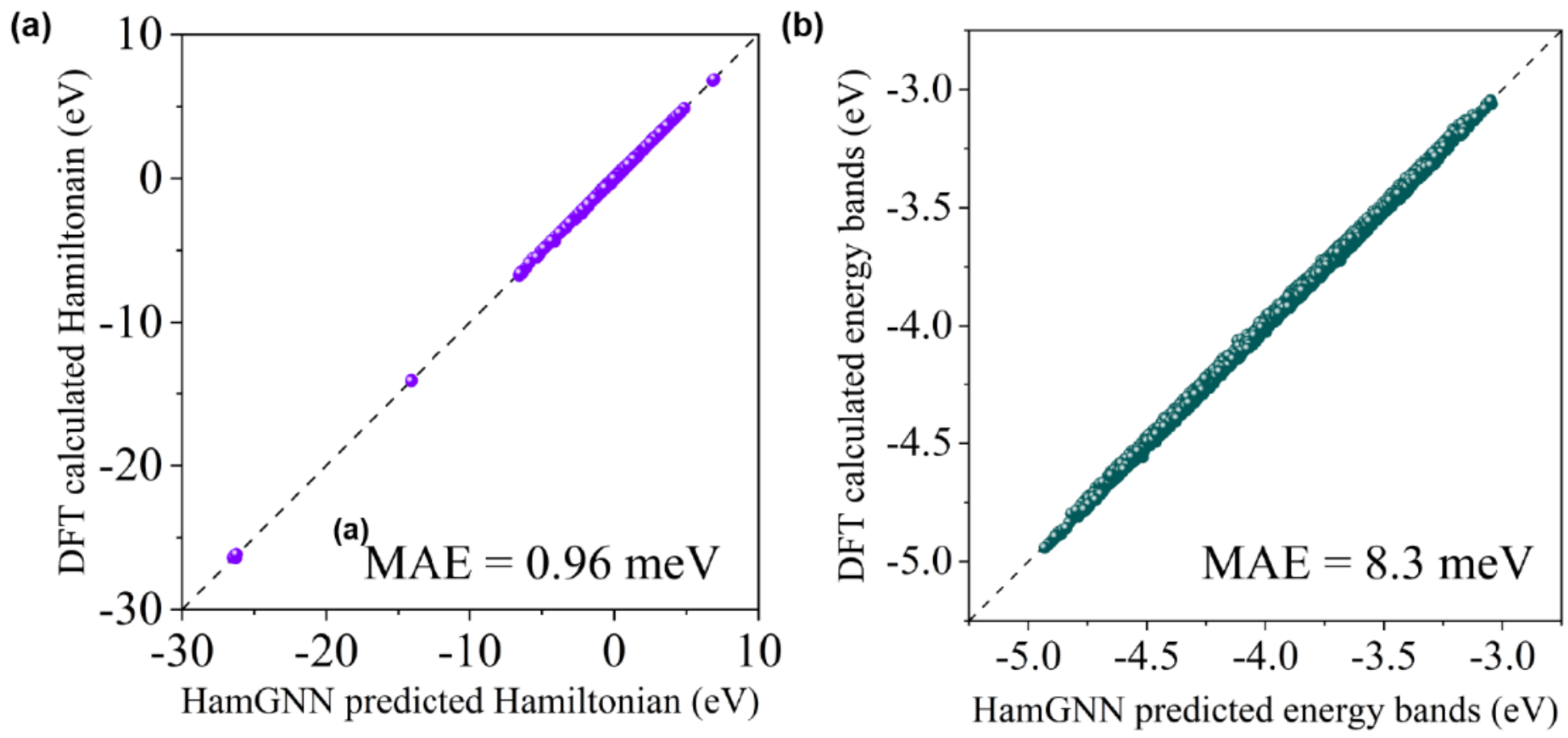

**FIG. S9**. **Demonstration of the accuracy of HamGNN.** (a) Comparison of the Hamiltonian matrix elements calculated by HamGNN and OpenMX on a test set. (b) Comparison of the energy bands near the Fermi levels calculated by HamGNN and OpenMX.

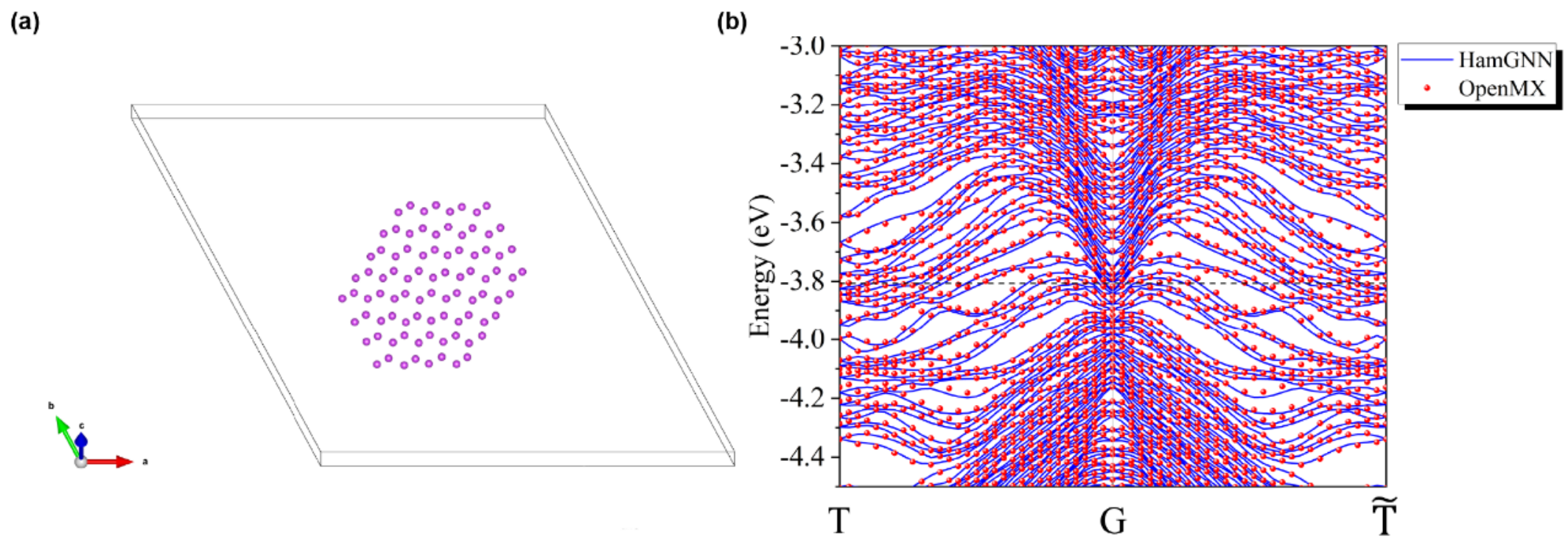

**FIG. S10**. **Bi prism test set.** (a) The crystal structure of a Bi prism in the test set. (b) Comparison of the energy bands calculated by HamGNN and OpenMX for the Bi prism shown in (a).

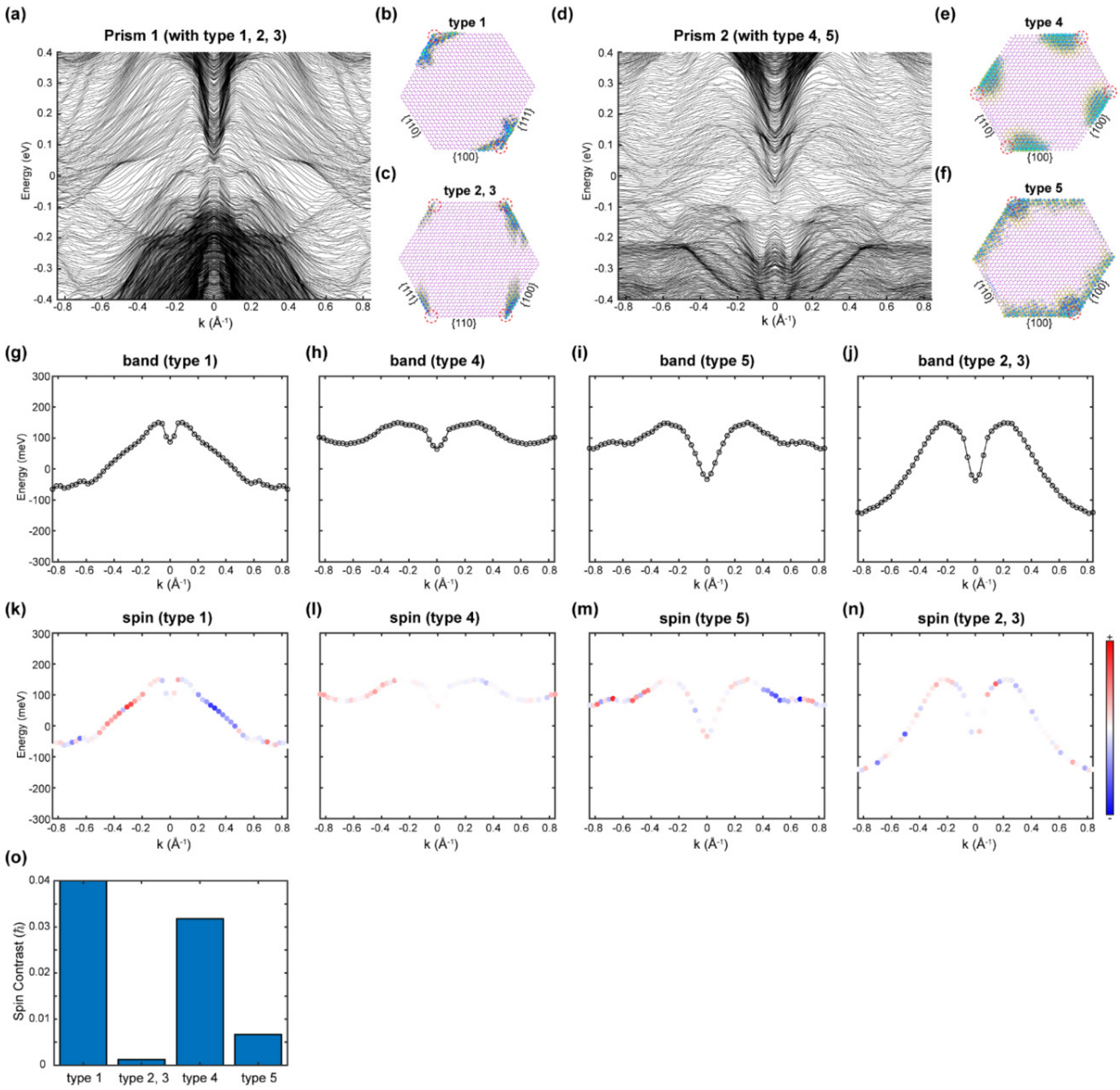

**FIG. S11**. **Calculation results of 5 types of hinges from two constructed prisms.** (a) and (d) Calculated band structure (including all the bulk and boundary states) of the two prisms along different directions, respectively. (b),(c),(e),(f) Real space distributions of the 1D hinge states whose wavefunctions are localized at different hinges. (g–j) E-k dispersion of the bands on all types of hinges. (k–n) Projected spin components of all types of hinge states. (o) Averaged spin polarization (contrast) of different hinge states.

## Part -IV: Symmetry analysis of the HOTI state in Bi

In this part we discuss the boundary states of bismuth follows the general approach described in refs. [9,10]. We first derive a Dirac-like Hamiltonian, which captures the topological nature of the bulk state of bismuth. We then study the boundary states of such Dirac-like Hamiltonian, by examining possible distributions of mass terms on the boundary. This discussion will reveal the existence of topologically protected hinge states in bismuth.

### III-1: Topological classification of bulk Bi

In this section, we aim to construct a Dirac-like Hamiltonian that matches the topological classification of bulk bismuth. The bulk bismuth has space-group symmetry $R\bar{3}m$, but since we only consider strong topological insulator, we ignore the translation symmetries and only consider its point-group symmetry $D_{3d}$. The full classification of free-fermion topological states with $D_{3d}$ symmetry is $\mathbb{Z} \times \mathbb{Z} \times \mathbb{Z}$ [11]. The formulation in Ref. [11] also allows us to represent each state in this $\mathbb{Z} \times \mathbb{Z} \times \mathbb{Z}$ classification by a Dirac-like Hamiltonian. Therefore, to construct a Dirac-like Hamiltonian, we need to determine the topological invariants of bulk bismuth in this $\mathbb{Z} \times \mathbb{Z} \times \mathbb{Z}$ classification.

The topological nature of bismuth has been studied in previous works [4,12], which determined that it is indeed a high-order topological insulator (HOTI). It has been shown that topological crystalline insulators with $R\bar{3}m$ symmetry are classified as $\mathbb{Z} \times \mathbb{Z}_2$ [13], and are characterized by two topological invariants: a mirror Chern number as an integer and a $\mathbb{Z}_2$ invariant characterizing a rotational HOTI. For bismuth, the two invariants are 0 and 1, respectively. We note the difference between $\mathbb{Z} \times \mathbb{Z} \times \mathbb{Z}$ and $\mathbb{Z} \times \mathbb{Z}_2$ classifications is that the former includes all atomic insulators (topological states without any boundary states), which are quotiented out in the latter classification.

To justify $\mathbb{Z} \times \mathbb{Z}_2$ is indeed $\mathbb{Z} \times \mathbb{Z} \times \mathbb{Z}$ modulo atomic insulators, we first note that the atomic insulators are given by the subgroup generated by $(2,2,0)$ and $(1,1,1)$ in $\mathbb{Z} \times \mathbb{Z} \times \mathbb{Z}$ [14,15, 2]. After quotienting out the subgroup of atomic insulators, we get a quotient group $\mathbb{Z} \times \mathbb{Z}_2$, and the quotient map from $\mathbb{Z} \times \mathbb{Z} \times \mathbb{Z}$ to $\mathbb{Z} \times \mathbb{Z}_2$ is given by $(1,0,0) \longrightarrow (1,1)$, $(0,1,0) \longrightarrow (-1,0)$ and $(0,0,1) \longrightarrow (0,1)$ (Note that the quotient map is not unique, since the basis of $\mathbb{Z} \times \mathbb{Z}_2$ is not unique). In another way, we can calculate the mirror Chern number and the $\mathbb{Z}_2$ invariant of rotational HOTI of the Hamiltonians representing three generators of $\mathbb{Z} \times \mathbb{Z} \times \mathbb{Z}$. Then $(0,0,1)$ again maps to $(0,1)$. $(1,0,0)$ and $(0,1,0)$ have mirror Chern number $+1$ and $-1$ respectively, but there is an ambiguity of the $\mathbb{Z}_2$ invariant of HOTI. Note that the $\mathbb{Z}_2$ invariant of HOTI asserts the appearance of higher boundary gapless states only when the bulk is not a strong topological insulator, that is, when the mirror Chern number is even. Otherwise, the whole boundary is gapless and it is not meaningful to discuss the appearance of higher boundary gapless states. Since $(1,0,0)$ and $(0,1,0)$ have odd mirror Chern number, the ambiguity of $\mathbb{Z}_2$ invariant of HOTI doesn't lead to any physical contradiction. In fact, if we take $(0,1,0)$ to be the “reference state”, i.e. the $\mathbb{Z}_2$ invariant of $(0,1,0)$ is 0, then the $\mathbb{Z}_2$ invariant of $(1,0,0)$ must be 1 (It is also meaningful to take another choice.). In this way, two quotient maps coincide with each other. This justifies our statement.

Therefore, Dirac-like Hamiltonians with different $\mathbb{Z} \times \mathbb{Z} \times \mathbb{Z}$ invariants but the same $\mathbb{Z} \times \mathbb{Z}_2$ invariants are belong to the same TCI phase (modulo atomic insulators), and they will have the same boundary states. In the following, we argue that, according to the $\mathbb{Z} \times \mathbb{Z} \times \mathbb{Z}$ classification, bismuth most likely has invariants $(0,0,1)$, and we therefore use the corresponding Dirac-like Hamiltonian in the next section to discuss its boundary states. However, choosing a different set of invariants that differs from $(0,0,1)$ by an atomic insulator will not change our conclusion of boundary states.

In the following, we extract the invariant in the full classification $\mathbb{Z} \times \mathbb{Z} \times \mathbb{Z}$ from the Hamiltonian $H_1(k) \oplus H_3(k)$ proposed in ref. [4] which captures the topology of Bismuth. The argument consists of two parts: (1) By ignoring all symmetries except $m_x$, $\boldsymbol{I}$ and $T$, we will get an invariant in $\mathbb{Z} \times \mathbb{Z}$ which is the full classification with symmetry $m_x$, $\boldsymbol{I}$ and $T$. The point group generated by $m_x$ and $\boldsymbol{I}$ is $C_{2h}$ with rotation axis $x$. (2) We construct a map from the full classfication $\mathbb{Z} \times \mathbb{Z}$ with symmetry $S_6$ to the Fu-Kane like invariant $\mathbb{Z}_2 \times \mathbb{Z}_2$.

$S_6$ is a subgroup of $D_{3d}$, so by forgetting the reflection $m_x$, there is a map from invariant of $D_{3d}$ to invariant of $S_6$. Thus, we construct a map from the full classification of $D_{3d}$ to the Fu-Kane like invariant. The Fu-Kane like invariant of $H_1(k) \oplus H_3(k)$ is known, it gives some information of the full classification through the map we constructed. After combining these two incomplete results, we will get the invariant in $\mathbb{Z} \times \mathbb{Z} \times \mathbb{Z}$.

The Hamiltonian of Bismuth is a direct sum of $H_j(k)$[4]

$$H_j(k) = (1 - k^2) 1 \otimes \sigma_z + \rho \cos(j\theta)\, \sigma_z \otimes \sigma_y + k_3 \sigma_x \otimes \sigma_x, \qquad (1)$$

where $k^2 = {k_1}^2 + {k_2}^2 + {k_3}^2$, $\rho$and $\theta$ are polar coordinate on the $k_1k_2$-plane, 1 and $\sigma_i$ are 2-dimensional identity matrix and Pauli matrices. $H_1(k)$ and $H_3(k)$ are $H_j(k)$ with $j = 1$ or 3. Note that $H_1(k)$ is a Dirac Hamiltonian and $H_3(k)$ is a "pasting" of three $H_1(k)$ since there is a 3 in front of $\theta$. The symmetry of $H_j(k)$ is given by $m_x = i\sigma_x \otimes 1$, $T = \sigma_y \otimes 1 \cdot K$, $S_6 = \mathrm{diag}\left\{e^{-i\frac{2\pi}{3}}, -e^{i\frac{2\pi}{3}}, e^{i\frac{2\pi}{3}}, -e^{-i\frac{2\pi}{3}}\right\}$ for $j = 1$ and $S_6 \otimes \sigma_z$ for $j = 3$.

The Hamiltonian is slightly different from the Hamiltonian in [4]. It is $\rho^j$ but not $\rho$ in the second and third terms. The original Hamiltonian is expanded from a tight binding Hamiltonian around $k = 0$. In the region closing to $k = 0$, our Hamiltonian is almost the same as the original Hamiltonian, but with a slower speed of changing of $\rho$ on the $k_1k_2$-plane. But when $k \to \infty$, our Hamiltonian is effectively $-1 \otimes \sigma_z$ from the $k^2$ term, but the original Hamiltonian is not well defined. The topology of a Dirac Hamiltonian is actually the topology of the Hamiltonian on the one-point compactification of the $k$-space, so it is not meaningful if the Hamiltonian is not well defined at infinity. To solve this defect, we should note that the $k^2$ term actually comes from the cosines in the tight binding Hamiltonian, but the $\rho^j$ term with $j = 1$ or 3 comes from the sines. The infinity of $k$-space corresponds to the boundary of the Brillouin zone, so it should be dominated by cosines but not sines. But if we expand the Hamiltonian around $k = 0$, the dominant term at infinity actually depends on the order of series we expand. By lowering the power of $\rho$, we make the infinity dominated by cosines again, and the speed of changing will not affect the topology of the Hamiltonian.

$H_1(k)$ is a Dirac Hamiltonian, according to [11], a Dirac Hamiltonian has the form of

$$H(k) = (1 - k^2) iM + \sum_{n=1}^{d} k_n \bar{J}_n. \qquad (2)$$

If we put $M$, $\bar{J}_i$, $i$, $T$, $S_6$and $m_x$ together, it will form a real $Cl^{3,3} \widehat{\otimes} \mathbb{R}[G]$-module by $\hat{S}_6 = \tilde{J}_3 e^{-\frac{\pi}{6}\tilde{J}_1\tilde{J}_2} S_6$, $\widehat{m}_x = \tilde{J}_1 m_x$, $J_1 = -iT$, $J_2 = T$ and $J_3 = M$. $Cl^{3,3}$ is the real Clifford algebra with generators $J_i$ and $\tilde{J}_i$, $\mathbb{R}[G]$ is a real group algebra, $\widehat{\otimes}$ represents the anticommutativity between generators of Clifford algebra and group elements that reverse the orientation. Notice that the imaginary unit $i$ is considered as a real matrix, and $K$ is considered as a real matrix anticommuting with $i$. There are six different nonisomorphic simple modules which corresponds to three generators and their inverses of $\mathbb{Z} \otimes \mathbb{Z} \otimes \mathbb{Z}$. The first and second generators correspond to a simple module with real dimension 8, and the third corresponds to a simple module with real dimension 16. To distinguish the first and the second, we can calculate $\hat{C}_2 = \hat{S}_6 \widehat{m}_x = 1$ or $-1$. By direct calculation, we will find that $H_1(k)$ correspond to the first generator.

We remain to determine the invariant of $H_3(k)$. The difficulty comes from that $H_3(k)$ is not a Dirac Hamiltonian, but we may expect $H_3(k)$ can be continuously transformed to a Dirac Hamiltonian in a symmetric way without closing the gap. Unfortunately, it is not easy to find such a transformation. But if we forget all symmetries except $m_x$, $I$and $T$, we will prove that $H_3(k)$ can be transformed to $H_1(k) \oplus H_1(-k_1, -k_2, k_3) \oplus H_1(-k_1, -k_2, k_3)$ under the symmetry $m_x$, $I$ and $T$. $m_x$ acts on each direct sum component in the same way as $m_x$ acting on $H_1(k)$. $I$ and $T$ not only act on each component, but also exchange the second and the third components using $\sigma_x$.

We divide the $k$-space to three regions, region A is $\frac{5\pi}{6} \le \theta \le \frac{7\pi}{6}$ and $-\frac{\pi}{6} \le \theta \le \frac{\pi}{6}$, region B is $\frac{\pi}{6} \le \theta \le \frac{5\pi}{6}$, and region C is $\frac{7\pi}{6} \le \theta \le \frac{11\pi}{6}$. We direct sum $H_3(k)$ with a trivial Hamiltonian $H_B(k) \oplus H_C(k)$ which will not change the topology of $H_3(k)$, we have

$$H_B(k) = H_3\left(\rho \cos\left(\frac{\pi}{6}\right), \rho \sin\left(\frac{\pi}{6}\right), k_3\right)$$

$$H_C(k) = H_3\left(\rho \cos\left(-\frac{\pi}{6}\right), \rho \sin\left(-\frac{\pi}{6}\right), k_3\right)$$

The symmetry $m_x$ is the same as $H_3(k)$ on $H_B(k)$ and $H_C(k)$, and $I$ and $T$ are also the same but followed by $\sigma_x \otimes 1 \otimes 1$ exchanging two Hamiltonians. Notice that $H_B(k)$ and $H_C(k)$ are constant under the change in $\theta$. They look like "flatten balloons", and there is an obvious contraction to a single point, so the topology is trivial. There is a continuous transformation $\mathcal{H}(k,t)$ from $H_3(k) \oplus H_B(k) \oplus H_C(k)$ to $H'_A(k) \oplus H'_B(k) \oplus H'_C(k)$

$$\mathcal{H}(k,t) = \begin{cases} H_3(k) \oplus H_B(k) \oplus H_C(k) & \text{region A} \\ U_B(t)\big(H_3(k) \oplus H_B(k) \oplus H_C(k)\big){U_B}^{-1}(t) & \text{region B}. \\ U_C(t)\big(H_3(k) \oplus H_B(k) \oplus H_C(k)\big){U_C}^{-1}(t) & \text{region C} \end{cases} \quad (4)$$

where

$$U_B(t) = \begin{pmatrix} \cos\left(\frac{\pi t}{2}\right) \cdot 1 & -\sin\left(\frac{\pi t}{2}\right) \cdot 1 & 0 \\ \sin\left(\frac{\pi t}{2}\right) \cdot 1 & \cos\left(\frac{\pi t}{2}\right) \cdot 1 & 0 \\ 0 & 0 & 1 \end{pmatrix}$$

$$U_C(t) = \begin{pmatrix} \cos\left(\frac{\pi t}{2}\right) \cdot 1 & 0 & -\sin\left(\frac{\pi t}{2}\right) \cdot 1 \\ 0 & 1 & 0 \\ \sin\left(\frac{\pi t}{2}\right) \cdot 1 & 0 & \cos\left(\frac{\pi t}{2}\right) \cdot 1 \end{pmatrix}. \quad (5)$$

It is straightforward to check that $\mathcal{H}(k,t)$ is symmetric. It should be careful that $\mathcal{H}(k,t)$ becomes $\mathcal{H}(-k,t)$, under the action $I$ and $T$ such that region B and C are exchanged. The result of the continuous transformation is $\mathcal{H}(k,1) = H'_A(k) \oplus H'_B(k) \oplus H'_C(k)$ with

$$H'_A(k) = \begin{cases} H_3(k) & \text{region A} \\ H_3\left(\rho \cos\left(\frac{\pi}{6}\right), \rho \sin\left(\frac{\pi}{6}\right), k_3\right) & \text{region B} \\ H_3\left(\rho \cos\left(-\frac{\pi}{6}\right), \rho \sin\left(-\frac{\pi}{6}\right), k_3\right) & \text{region C} \end{cases}$$

$$H'_B(k) = \begin{cases} H_3(k) & \text{region B} \\ H_3\left(\rho\cos\left(\frac{\pi}{6}\right), \rho\sin\left(\frac{\pi}{6}\right), k_3\right) & \text{region A and C} \end{cases}$$

$$H'_C(k) = \begin{cases} H_3(k) & \text{region C} \\ H_3\left(\rho\cos\left(-\frac{\pi}{6}\right), \rho\sin\left(-\frac{\pi}{6}\right), k_3\right) & \text{region A and B} \end{cases}. \quad (6)$$

$H'_A(k)$, $H'_B(k)$ and $H'_C(k)$ have two regions to be constant in $\theta$, so there is a continuous transformation to shrink these constant regions. The result of shrinking is $H_1(k)$, $H_1(-k_1,-k_2,k_3)$ and $H_1(-k_1,-k_2,k_3)$. The symmetry $m_x$ is the same as $H_1(k)$, and $I$ and $T$ are the same as $H_1(k)$ but followed by exchanging of the last two Hamiltonians.

Now we have proved $H_1(k)$ has the same topology with a Dirac Hamiltonian $H_1(k) \oplus H_1(-k_1,-k_2,k_3) \oplus H_1(-k_1,-k_2,k_3)$. It remains to get the invariant of the Dirac Hamiltonian in $\mathbb{Z}\times\mathbb{Z}\times\mathbb{Z}$. Notice that $I$ and $T$ exchange the second and third components. There is an obvious change of basis such that the Hamiltonians are not changed, and $I$ and $T$ act on the second Hamiltonian in the same way as $H_1(k)$ but the third with a minus sign. Now the Hamiltonian is a direct sum of three independent Dirac Hamiltonians. By comparing these Hamiltonians to the Hamiltonians constructed from the classification $\mathbb{Z}\times\mathbb{Z}$ of topological insulator with symmetry $C_{2h}$ [11], we will get a invariant of $H_3(k)$ in $\mathbb{Z}\times\mathbb{Z}$.

In the classification $\mathbb{Z}\times\mathbb{Z}$ of $C_{2h}$, two generators are all simple modules of real dimension 8. To distinguish them, the first one has $C_2 = 1$, the second one has $C_2 = -1$.

We check the generators of three Dirac Hamiltonians as the algebra $Cl^{3,3} \widehat{\otimes} \mathbb{R}[C_{2h}]$. $C_{2h}$ is the point group generated by $m_x$ and$I$, since they reverse the orientation, $m_x$ and $I$ anticommute with Clifford generators. The rotation is $C_2 = m_x I$ along axis $x$. For the first Dirac Hamiltonian, we have $\widehat{m}_x = \tilde{J}_1 m_x$, $\hat{I} = \tilde{J}_1\tilde{J}_2\tilde{J}_3 I$, $J_1 = -iT$, $J_2 = T$, $J_3 = M$, $\tilde{J}_1$, $\tilde{J}_2$ and $\tilde{J}_3$. For the second Dirac Hamiltonian, we have $\widehat{m}_x = -\tilde{J}_1 m_x$, $\hat{I} = \tilde{J}_1\tilde{J}_2\tilde{J}_3 I$, $J_1 = -iT$, $J_2 = T$, $J_3 = M$, $-\tilde{J}_1$, $-\tilde{J}_2$ and $-\tilde{J}_3$. For the third Dirac Hamiltonian, we have $\widehat{m}_x = -\tilde{J}_1 m_x$, $\hat{I} = -\tilde{J}_1\tilde{J}_2\tilde{J}_3 I$, $J_1 = iT$, $J_2 = T$, $J_3 = M$, $-\tilde{J}_1$, $-\tilde{J}_2$ and $\tilde{J}_3$.

The first Hamiltonian is the same as $H_1(k)$ with $C_2 = 1$, so we set it as $(1,0) \in \mathbb{Z}\times\mathbb{Z}$. The second Hamiltonian has $C_2 = -1$, and it represent $(0,1) \in \mathbb{Z}$. The third Hamiltonian is the same the first, but with odd number of minus signs on those generators. Notice that $m_x$ and $I$ are the same, so it only counts once. These odd number of minus signs imply that the third Hamiltonian is the inverse of the first which is $(-1,0) \in \mathbb{Z}\times\mathbb{Z}$. In summary, $H_3(k)$ has the invariant $(0,1) \in \mathbb{Z}\times\mathbb{Z}$.

By ignoring other symmetries, there is a map from $\mathbb{Z}\times\mathbb{Z}\times\mathbb{Z}$ of $D_{3d}$ to $\mathbb{Z}\times\mathbb{Z}$ of $C_{2h}$, which is induced by $(1,0,0) \to (1,0)$, $(0,1,0) \to (0,1)$ and $(0,0,1) \to (1,1)$. We get to a conclusion that $H_3(k)$ has the invariant of $(a, 1+a, -a)$ where $a \in \mathbb{Z}$, since $(a, 1+a, -a)$ is the preimage of $(0,1) \in \mathbb{Z}\times\mathbb{Z}$.

It remains an integer $a$ to be determined. We first prove that $a$ is odd. It can be obtained from the Fu-Kane like invariant in [4]. The classification with $S_6$ is $\mathbb{Z}\times\mathbb{Z}$. $S_6$ is a subgroup of $D_{3d}$ by forgetting the reflection $m_x$, so it induces a map from invariant $\mathbb{Z}\times\mathbb{Z}\times\mathbb{Z}$ of $D_{3d}$ to invariant $\mathbb{Z}\times\mathbb{Z}$ of $S_6$. The map is spanned by $(1,0,0) \to (1,0)$, $(0,1,0) \to (0,1)$ and $(0,0,1) \to (1,1)$. The Fu-Kane like invariant of $H_3(k)$ is known to be $(1,0) \in \mathbb{Z}_2\times\mathbb{Z}_2$. We need to construct a map from classification $\mathbb{Z}\times\mathbb{Z}$ of $S_6$ to Fu-Kane like invariant $\mathbb{Z}_2\times\mathbb{Z}_2$. It can be obtained by calculating the Fu-Kane like invariant of two generators of full classification of $S_6$.

The Dirac Hamiltonian constructed from the invariant is defined on the one-point compactification of $k$-space. All time reversal symmetric points are on the boundary of Brillouin zone except the $\Gamma$ point. The infinity of $k$-space corresponds to the boundary of Brillouin zone, so the eigenvalues of $I$ of the occupied bands at the time reversal symmetric points are all the same except the one at $\Gamma$. It turns out that two Fu-Kane invariants $\nu^{(\pi)}$ and $\nu^{(\pm\pi/3)}$ are determined by the difference between eigenvalues of $I$ of the occupied bands at $k=0$ and infinity. The Dirac Hamiltonian from the first generator of $S_6$ in $\mathbb{Z}\times\mathbb{Z}$ is the same as $H_1(k)$, and it is easy to get $(0,1)\in\mathbb{Z}_2\times\mathbb{Z}_2$ The Hamiltonian from the second generator is the same as Eq. (7) in the following. Half of the eigenvalues of $C_3=-{S_6}^2$ is $-1$ and another half is $e^{\pm i\pi/3}$. The minus sign in front of ${S_6}^2$ comes from the spinful representaion, so the symmetry operator is only determined up to a minus sign. It is also easy to get $(1,1)\in\mathbb{Z}_2\times\mathbb{Z}_2$.

Since $H_3(k)$ has $(1,0)\in\mathbb{Z}_2\times\mathbb{Z}_2$ according to [4], $H_3(k)$ must be $(c,d)\in\mathbb{Z}\times\mathbb{Z}$ of $S_6$ with $c$ and $d$ odd. This further proves the third generators of $\mathbb{Z}\times\mathbb{Z}\times\mathbb{Z}$ of $D_{3d}$ must be odd. This completes the proof that $a$ is odd. Next, we conjecture that $a=-1$, based on the following argument. In fact, the difference of states with different $a$ is an atomic insulator, so the correctness of this conjecture will not affect our discussion of boundary states. Note that $H_3(k)$ is a "pasting"' of three Dirac Hamiltonians with real dimension 8, so it likely does not have the ability to support a complicated Hamiltonian with invariant adding to bigger than 3. Recall that the first and second invariant in $\mathbb{Z}\times\mathbb{Z}\times\mathbb{Z}$ has real dimension of 8, and the third has real dimension of 16 which has twice complication than the one with 8 dimensions, so any $a$ other than $0$ and $-1$ are too complicated.

In summary, $H_1(k)$ has $(1,0,0)\in\mathbb{Z}\times\mathbb{Z}\times\mathbb{Z}$, and $H_3(k)$ has $(-1,0,1)\in\mathbb{Z}\times\mathbb{Z}\times\mathbb{Z}$. Thus $H_1(k)\oplus H_3(k)$ has $(0,0,1)\in\mathbb{Z}\times\mathbb{Z}\times\mathbb{Z}$. Therefore, we will use the Dirac Hamiltonian constructed from the topological invariant [2] to study the boundary state of bismuth,

$$H(k)=iM+k^2 iM_0+\sum_{n=1}^{d}k_n\tilde{J}_n, \quad (7)$$

where

$$\tilde{J}_1=1\otimes 1\otimes\sigma_y$$

$$\tilde{J}_2=1\otimes 1\otimes\sigma_x$$
$$\tilde{J}_3=1\otimes\sigma_z\otimes\sigma_z$$
$$M=-M_0=1\otimes i\sigma_y\otimes\sigma_z$$
$$T=1\otimes 1\otimes i\sigma_y K$$
$$S_6=\tilde{J}_3 e^{\frac{\pi}{6}\tilde{J}_1\tilde{J}_2}\begin{pmatrix}\cos\left(\frac{4\pi}{3}\right) & -\sin\left(\frac{4\pi}{3}\right)\\ \sin\left(\frac{4\pi}{3}\right) & \cos\left(\frac{4\pi}{3}\right)\end{pmatrix}\otimes 1\otimes 1\cdot 1\otimes\sigma_x\otimes\sigma_z$$
$$m_x=\tilde{J}_1\sigma_z\otimes\sigma_x\otimes\sigma_z. \quad (8)$$

$S_6$ and $m_x$ are generators of spinful symmetry $D_{3d}$. $S_6$ rotate $\pi/3$ along axis $z$ and then reflect along axis $z$. $m_x$ is the reflection along axis $x$. They satisfy ${S_6}^6=1$, ${m_x}^2=-1$ and $(S_6 m_x)^2=-1$.

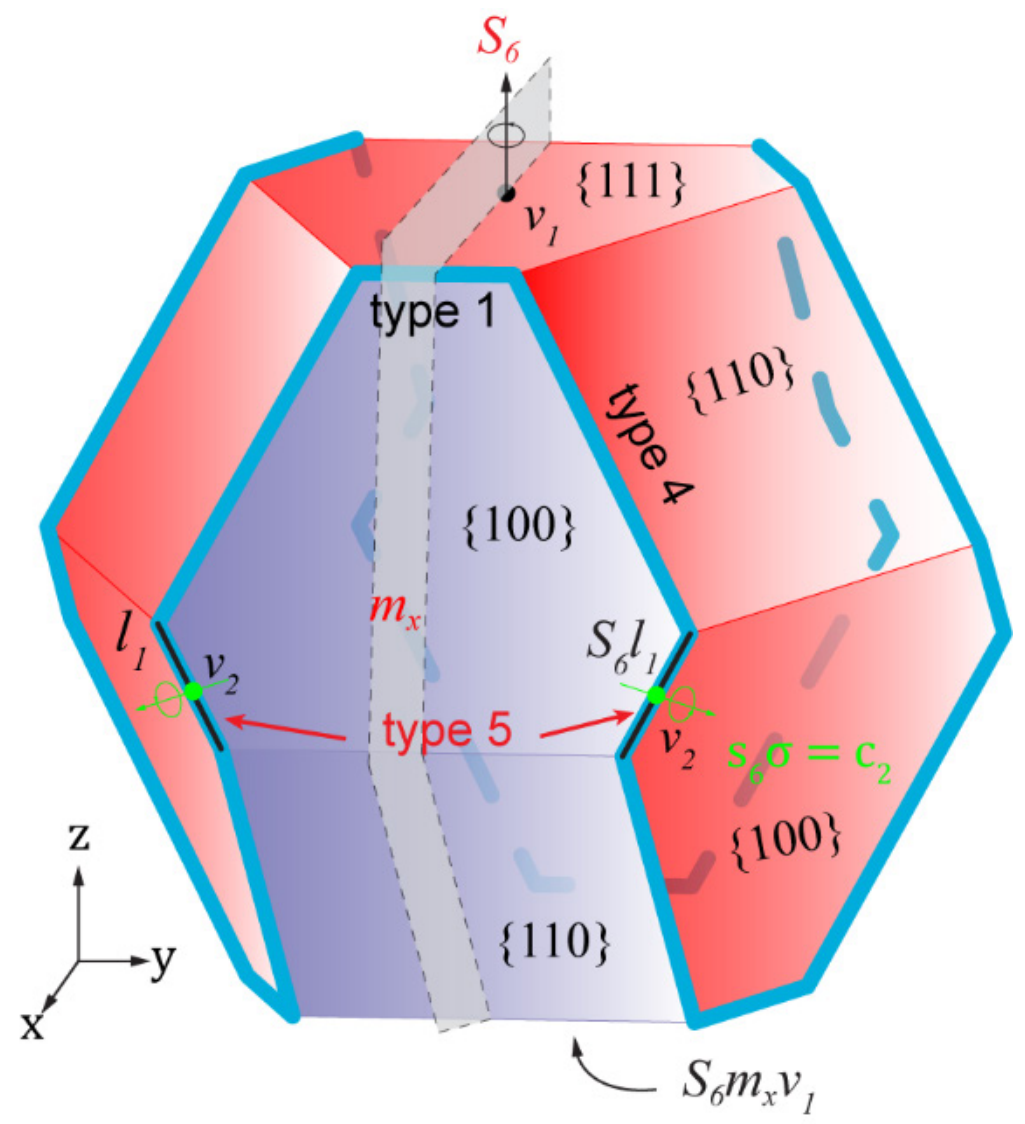

**FIG. S12.** Schematic of HOTI property in Bi nanocrystal and symmetry analysis. Two generators of the $D_{3d}$ point group of Bi are marked by red symbols. $S_6$ is the six-fold improper rotation and $m_x$ is the reflection, respectively. $v_1$, $v_2$ and $S_6 m_x v_1$ are high-symmetry points on top/bottom {111} facets and type 5 hinge.

## III-2: The boundary

In this section, we study the boundary of Bismuth, using the Dirac Hamiltonian in Eq. (7). The crystal model of Bi is now shown in Fig. S12. We will prove the following rules: 1) the vertex $v_2$ must be anomalous. 2) For any paths between $v_1$ and $S_6 m_x v_1$, the path will cross through an anomaly. 3) If the anomaly in 1) and 2) appears, all other regions can be gapped out. Here the anomaly means a certain region (a hinge or a corner) has crystalline-symmetry-protected gapless state, which is generally referred as topological hinge or corner state.

In Eq. (7), $M$ is the mass term. A mass term should satisfy $M^2 = -1$, $\{M, \tilde{J}_i\} = 0$, and $iM$ is invariant under the symmetry transformation. A boundary mass term $M'$ (a mass term that can gap out the boundary Hamiltonian) anti-commutes with the mass term $M$ of the Hamiltonian. To gap out the boundary of the crystalline grain, we need to find a gapped boundary term in a symmetric and continuous way on the whole surface of the crystalline grain. The meaning of continuity is that the boundary mass term $M'(x)$ is a continuous function which continuously depends on the parameter $x$ on the surface. The meaning of symmetric is that the boundary mass term satisfies $gM'(x)g^{-1} = M'(gx)$ for $g \in D_{3d}$.

Note that different points on the boundary may have different symmetry (which is described by the subgroup satisfying $gx = x$ for some point $x$ on the surface). We show that each point of the boundary has a boundary mass term except the vertex $v_2$ regardless of the continuity ($v_2$ is on the type 5 hinge). Consider the boundary mass term $M' = i\sigma_y \otimes \sigma_x \otimes \sigma_z$, $iM'$ is invariant under the action of ${S_6}^2$, $m_x$ and $T$. So, all points on the boundary had been gapped out except the orbit of $v_2$. In fact, it is impossible to find a boundary mass term which is invariant under ${S_6}^{-1} m_x$, according to the corresponding classification result in [11]. This proves rule 1).

Although all points can be gapped out except $v_2$, it doesn't mean that we can gap out the whole boundary except $v_2$, because the boundary mass term we choose should also be symmetric and continuous. To prove rule 2), we note that if the boundary mass term on $v_1$ is taken to be $M'$, then the requirement of symmetry demands the boundary mass term on $S_6 m_x v_1$ to be $S_6 m_x M'^{(S_6 m_x)^{-1}} = -M'$. We need to find a continuous transformation between $M'$ and $-M'$ because of the requirement of continuity. It turns out that we need to find one more boundary mass term $M''$which anticommutes with both $M$ and $M'$. Again, it can be proved that this is also impossible [9,10]. This proves 2). Along with 1), 3) are also proved.

Since type-5 hinge contains the high-symmetry point $v_2$ and it is symmetric under $s_6$ and $m_x$, it must host gapless states. Rule 2) means the top and bottom $\{111\}$ facts belong to two different regions separated by a loop of gapless state; and the two type-5 hinges, like the edges $l_1$ and $s_6 l_1$, must be connected by a gapless hinge state. According to the above analysis and the result of our experiment, the only possibility of boundary states is shown in Fig. 4a. The topological hinge states in the type 1,4,5 hinges form a closed loop and divided the facets in to two separate regions. These two surface regions have mass terms of $\pm M'$, respectively. Therefore, gapless hinge states will emerge at their intersections. This is a direct consequence of the bulk-boundary correspondence of a HOTI.

References

[1]. Guo, Z., Deng, J., Xie, Y. & Wang, Z. Quadrupole topological insulators in Ta2M3Te5 (M = Ni, Pd) monolayers. *npj Quantum Mater*. 7, 87 (2022).

[2]. Yue, C. et al. Symmetry-enforced chiral hinge states and surface quantum anomalous Hall effect in the magnetic axion insulator $Bi_{2-x}Sm_xSe_3$. *Nat. Phys.* 15, 577-581 (2019).

[3]. Ren, Y., Qiao, Z. & Niu, Q. Engineering Corner States from Two-Dimensional Topological Insulators. *Phys. Rev. Lett.* 124, 166804 (2020).

[4]. Schindler, F. et al. Higher-Order Topology in Bismuth. *Nat. Phys.* 14, 918-924 (2018).

[5]. Bernevig, B. A., Hughes, T. L. & Zhang, S.-C. Quantum Spin Hall Effect and Topological Phase Transition in HgTe Quantum Wells. *Science* 314, 1757-1761 (2006).

[6]. Zhong, Y., Yu, H., Su, M., Gong, X. & Xiang, H. Transferable equivariant graph neural networks for the Hamiltonians of molecules and solids. *npj Comput. Mater.* 9, 182 (2023).

[7]. Ozaki, T. Variationally optimized atomic orbitals for large-scale electronic structures. *Phys. Rev. B* 67, 155108 (2003).

[8]. Ozaki, T. & Kino, H. Numerical atomic basis orbitals from H to Kr. *Phys. Rev. B* 69, 195113 (2004).

[9]. Michael Stone, Ching-Kai Chiu, and Abhishek Roy. Symmetries, dimensions and topological insulators: the mechanism behind the face of the bott clock. *Journal of Physics A: Mathematical and Theoretical*, 44(4):045001, dec 2010.

[10]. Tian Yuan, Chen Fang, and Yang Qi. In preparation.

[11]. Eyal Cornfeld and Adam Chapman. Classification of crystalline topological insulators and superconductors with point group symmetries. *Phys. Rev. B*, 99, 075105 (2019).

[12]. Chuang-Han Hsu, Xiaoting Zhou, Tay-Rong Chang, Qiong Ma, Nuh Gedik, Arun Bansil, Su-Yang Xu, Hsin Lin, and Liang Fu. Topology on anew facet of bismuth. Proceedings of the National *Academy of Sciences*,116, 13255-13259 (2019).

[13]. Zhida Song, Sheng-Jie Huang, Yang Qi, Chen Fang, and Michael Hermele. Topological states from topological crystals. *Science Advances*, 5, eaax2007 (2019).

[14]. Eyal Cornfeld and Shachar Carmeli. Tenfold topology of crystals: Unified classification of crystalline topological insulators and superconductors. *Physical Review Research*, 3(1):013052 (2021).

[15]. Ken Shiozaki. The classification of surface states of topological insulators and superconductors with magnetic point group symmetry. *Progress of Theoretical and Experimental Physics*, 2022(4):04A104, (2022).